\documentclass[final,3p,times,sort&compress]{elsarticle}

\usepackage{amssymb,amsmath}

\usepackage{bm,setspace}
\usepackage[electronic]{ifsym}

\newcommand{\rd}{\mathrm{d}}
\newcommand{\e}{\mathrm{e}}
\newcommand{\ri}{\mathrm{i}}
\newcommand{\ep}{\varepsilon}
\newcommand{\vx}{\bm{x}}
\newcommand{\vv}{\bm{\varv}}
\newcommand{\vnabla}{\bm{\nabla}}
\newcommand{\stepup}{\text{\RaisingEdge}}
\newcommand{\stepdown}{\text{\,\FallingEdge}}
\newcommand{\hres}{h_\mathrm{res}}
\newcommand{\tod}{\stackrel{\mathrm{d}}{\to}}
\newcommand{\unit}[1]{~\mathrm{#1}}
\renewcommand{\(}{\left(}
\renewcommand{\)}{\right)}
\renewcommand{\[}{\left[}
\renewcommand{\]}{\right]}
\DeclareMathOperator{\im}{Im}
\DeclareMathOperator{\re}{Re}
\DeclareMathOperator{\Tr}{Tr}
\DeclareMathOperator{\Prob}{Prob}
\DeclareMathOperator{\Ai}{Ai}
\DeclareMathOperator{\Sk}{Sk}
\DeclareMathOperator{\Ku}{Ku}
\newcommand{\const}{\text{const.}}
\newcommand{\diff}[2]{\frac{\mathrm{d} #1}{\mathrm{d} #2}}
\newcommand{\diffs}[3]{\frac{\mathrm{d}^{#3} #1}{\mathrm{d}{#2}^{#3}}}
\newcommand{\prt}[2]{\frac{\partial #1}{\partial #2}}
\newcommand{\prts}[3]{\frac{\partial^{#3} #1}{\partial {#2}^{#3}}}
\newcommand{\varder}[2]{\frac{\delta #1}{\delta #2}}
\newcommand{\varders}[3]{\frac{\delta^{#3} #1}{\delta^{#3} #2}}
\newcommand{\expct}[1]{\langle #1 \rangle}
\newcommand{\Expct}[1]{\left\langle #1 \right\rangle}
\newcommand{\cum}[1]{\langle #1 \rangle_\mathrm{c}}

\newcommand{\floor}[1]{\lfloor #1 \rfloor}
\newcommand{\ket}[1]{| #1 \rangle}   
\newcommand{\braket}[3]{\langle #1 | #2 | #3 \rangle}
\newcommand{\etal}{\textit{et al.}}
\newcommand{\figref}[1]{Fig.~\ref{#1}}
\newcommand{\tblref}[1]{Table~\ref{#1}}
\newcommand{\secref}[1]{Sec.~\ref{#1}}
\newcommand{\pref}[1]{(\ref{#1})}
\renewcommand{\eqref}[1]{Eq.~\pref{#1}}

\journal{Physica A}

\begin{document}

\begin{frontmatter}



\title{An appetizer to modern developments on the Kardar-Parisi-Zhang universality class}


\author{Kazumasa A. Takeuchi}

\address{Department of Physics, Tokyo Institute of Technology,\\
 2-12-1 Ookayama, Meguro-ku, Tokyo 152-8551, Japan.}

\begin{abstract}
The Kardar-Parisi-Zhang (KPZ) universality class describes
 a broad range of non-equilibrium fluctuations,
 including those of growing interfaces, directed polymers
 and particle transport, to name but a few.
Since the year 2000, our understanding of the one-dimensional KPZ class
 has been completely renewed by mathematical physics approaches
 based on exact solutions.
Mathematical physics has played a central role since then,
 leading to a myriad of new developments,
 but their implications are clearly not limited to mathematics
 -- as a matter of fact, it can also be studied experimentally.
The aim of these lecture notes is to provide an introduction
 to the field that is accessible to non-specialists, 
 reviewing basic properties of the KPZ class
 and highlighting main physical outcomes of mathematical developments
 since the year 2000.
It is written in a brief and self-contained manner,
 with emphasis put on physical intuitions and implications,
 while only a small (and mostly not the latest) fraction
 of mathematical developments could be covered.
Liquid-crystal experiments by the author and coworkers
 are also reviewed.

\end{abstract}


\end{frontmatter}


\section{Introduction}
\label{sec:Introduction}

Physics of critical phenomena, as represented by those of the Ising model,
 is one of the monuments of statistical physics.
From the author's possibly biased view,
 it was founded cooperatively by theoretical studies
 \cite{Nishimori.Ortiz-Book2011},
 both rigorous
 and non-rigorous,
 as well as by experiments \cite{Heller-RPP1967}.
More precisely, when Andrews discovered
 the liquid-vapor critical point in the 19th century \cite{Andrews-PTRSL1869},
 what he observed was the Ising critical behavior from the modern viewpoint,
 though the Ising model was invented much later.
In the mid 20th century,
 the two-dimensional Ising model was solved exactly by Onsager,
 and subsequently by Kaufman and Nambu
 \cite{Baxter-Book1982,Onsager-PR1944,Kaufman-PR1949,Nambu-PTP1950}.
This exact solution clearly indicated
 the existence of nontrivial critical behavior
 that is different from the prediction of the mean field theory.
Moreover, it turned out that critical phenomena exhibit
 certain extent of universality, as suggested by experimental observations
 of liquid-vapor systems,
 binary liquid mixtures and magnets \cite{Heller-RPP1967}.
This universality was clearly accounted for by Wilson's renormalization group
 \cite{Wilson-RMP1975} on the basis of continuum equations
 such as the $\phi^4$ model, which allowed us to classify critical phenomena
 in terms of universality classes.
This line of research marked further milestones,
 such as the foundation of conformal field theory \cite{Henkel-Book1999},
 and it continues doing so,
 as exemplified by recent developments of conformal bootstrap theory
 \cite{El-Showk.etal-JSP2014} tackling the three-dimensional Ising problem.

All these developments concern critical phenomena at thermal equilibrium.
Then, does there exist such a beautiful physics
 for systems driven out of equilibrium?
No one knows the answer yet, but developments that might resemble
 the dawn of the equilibrium counterpart can be found in recent studies
 on the Kardar-Parisi-Zhang (KPZ) universality class
 \cite{Kardar.etal-PRL1986,Barabasi.Stanley-Book1995,HalpinHealy.Zhang-PR1995,Kriecherbauer.Krug-JPA2010,Corwin-RMTA2012,HalpinHealy.Takeuchi-JSP2015,Sasamoto-PTEP2016}.
The KPZ class, best known as the simplest generic class
 for fluctuations of growing interfaces \cite{Kardar.etal-PRL1986},
 is also related to a wide variety of non-equilibrium fluctuations,
 including directed polymers, stirred fluids, and particle transport
 to name but a few.
Furthermore, for the one-dimensional (1D) case,
 mathematical studies have unveiled
 nontrivial connections to random matrix theory, combinatorial problems,
 and integrable systems.
This has driven intense activities to investigate exact fluctuation properties
 of the 1D KPZ class and its mathematical structure behind,
 providing a great number of exact results
 \cite{Kriecherbauer.Krug-JPA2010,Corwin-RMTA2012,HalpinHealy.Takeuchi-JSP2015,Quastel.Spohn-JSP2015,Sasamoto-PTEP2016}
 which also have experimental relevance
 \cite{Takeuchi.Sano-PRL2010,Takeuchi.etal-SR2011,Takeuchi.Sano-JSP2012}.

The aim of these lecture notes is to tempt non-specialists
 into this rapidly evolving field around the KPZ class.
It is \textit{not} a review of recent mathematical approaches
 nor a technical guide to solve the problems,
 for which the readers are referred to more appropriate reviews \cite{Kriecherbauer.Krug-JPA2010,Corwin-RMTA2012,Quastel.Spohn-JSP2015,Sasamoto-PTEP2016}
 and references therein.
Instead, it is intended to provide useful information
 for non-specialist physicists to join the game,
 in a brief but self-contained manner:
 how the KPZ class is linked to various problems,
 what the main outcomes and implications for physicists are,
 and what kind of intuitions we can use,
 from the limited view of an experimentalist admirer and user
 of those mathematical developments.

These lecture notes are organized as follows.
Section \ref{sec:Interface} describes what kind of interfaces
 we deal with, on the basis of some experimental observations.
Section \ref{sec:Continuum} introduces continuum equations
 and their associated universality classes,
 including the KPZ equation and the KPZ class,
 and review their basic properties.
Properties of the KPZ equation are further described in \secref{sec:KPZeq}.
Sections \ref{sec:Exact1} and \ref{sec:Exact2} illustrate
 some exact results for the 1D KPZ class,
 focusing on the distribution of interface fluctuations.
Section \ref{sec:Exp} reviews experimental observations
 of growing interfaces in turbulent liquid crystal,
 to be compared with the exact results described in the preceding sections.
In \secref{sec:HigherDimension}
 we briefly discuss the situation in higher dimensions.
Section \ref{sec:Perspectives}
 provides brief concluding remarks.

\section{Examples of quiescent and growing interfaces}  \label{sec:Interface}

The surface of water at rest is a symbol of something flat and smooth.
Indeed, according to the capillary wave theory \cite{Rowlinson.Widom-Book2002},
 which accounts for thermal excitations
 of such a free interface between two fluid phases,
 the amplitude of fluctuations under gravity
 is in the order of $\sqrt{k_\mathrm{B}T/\gamma}$
 with the Boltzmann constant $k_\mathrm{B}$, 
 temperature $T$, and interfacial tension $\gamma$.
For molecular fluids, it is usually shorter than 1~nm
 \cite{Rowlinson.Widom-Book2002,Aarts.etal-S2004};
 therefore, the interface is sufficiently smooth
 when observed at macroscopic length scales.
But what happens if one of the phases is more stable than the other,
 hence taking over the region of the metastable state?
What happens if one of the phases is solid and molecules are deposited
 on it one after another, such as in thin film growth?
What happens if there is an aggregate of particles that can replicate
 themselves, such as living cells?
Or, conversely, what happens if particles are being removed
 from the surface of the aggregate?
In all such cases the interface (or the edge of the aggregate)
 will move in either direction, typically with fluctuations
 growing in time.

\begin{figure}[t]
 \centering
 \includegraphics[scale=0.8,clip]{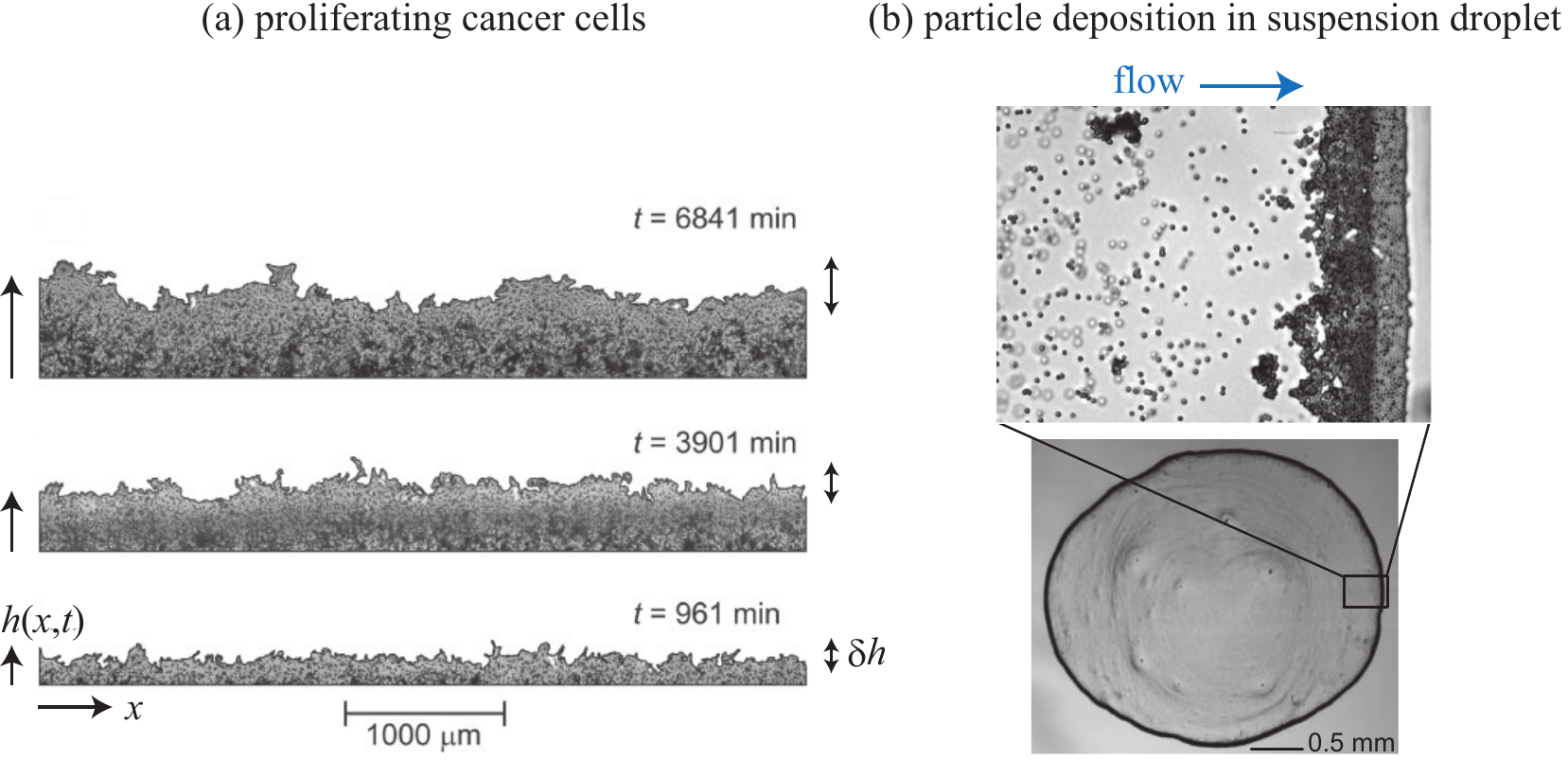}
 \caption{
Experimental examples of growing interfaces.
(a) Front evolution of cancer cell colonies
 (``HeLa cells'' from human cervix cancer)
 cultured in Petri dishes \cite{Huergo.etal-PRE2012}.
Here the colonies are two-dimensional,
 so that the interfaces are one-dimensional,
 and they grow because of the division of cancer cells.
The 1D KPZ exponents were found \cite{Huergo.etal-PRE2012}.
Similar growth was also observed with Vero cells
 (from African Green Monkey kidney), which are not tumorigenic but invasive
 \cite{Huergo.etal-PRE2010,Huergo.etal-PRE2011}.
(b) Deposition of spherical colloid particles
 onto the edge of a suspension droplet, during its evaporation process
 \cite{Yunker.etal-N2011}.
Fluid flows outward because of capillary effect, carrying colloids
 and leaving them at the droplet edge.
The 1D KPZ exponents were observed
 when slightly elongated particles were used \cite{Yunker.etal-PRL2013}.
Reprints with permission from
 \cite{Huergo.etal-PRE2012}(a) and \cite{Yunker.etal-N2011}(b)
 with some adaptations.
The axes and the arrows, as well as the associated labels,
 are added by the present author.
The displayed amplitudes of $\delta h$ are only schematic.
}
 \label{fig1}
\end{figure}%

Let's see some examples.
Figure~\ref{fig1}(a) shows snapshots of colonies
 of human cancer cells (HeLa), cultured in Petri dishes
 \cite{Huergo.etal-PRE2012}.
Those cells were initially cultured with an obstacle attached
 on the Petri dish bottom, which set the initially straight border of colonies.
When the obstacle was removed, the colony starts to expand
 because of cell division and motility.
As a result, the interface moves in the normal direction,
 but fluctuations also start to develop, because of the stochastic nature
 of cell behavior.
Then the snapshots in \figref{fig1}(a) suggest that
 those interface fluctuations gain larger and larger structure as time elapses.

Figure~\ref{fig1}(b) shows another example, taken from an experiment
 of ``coffee ring effect'' \cite{Yunker.etal-N2011}
 -- a familiar phenomenon observed
 when a droplet of coffee is dried on a solid surface,
 which leaves a ring of coffee particles on the surface.
The formation of the ring is due to the deposition of particles
 onto the droplet edge, carried by fluid capillary flow.
In \figref{fig1}(b), polystyrene beads were used as particles,
 and the deposition was monitored by a microscope
 (see also Supplementary Movie 3 of \cite{Yunker.etal-N2011}).
One can again see that, as the front of the deposited particles moves inward,
 its fluctuations develop, because of random deposition of particles.
Here we can also see that those particles interact;
 sometimes they collide and stick to each other, and in fact they can interact
 through deformation of the air-liquid surface.

\begin{figure}[t]
 \centering
 \includegraphics[scale=0.8,clip]{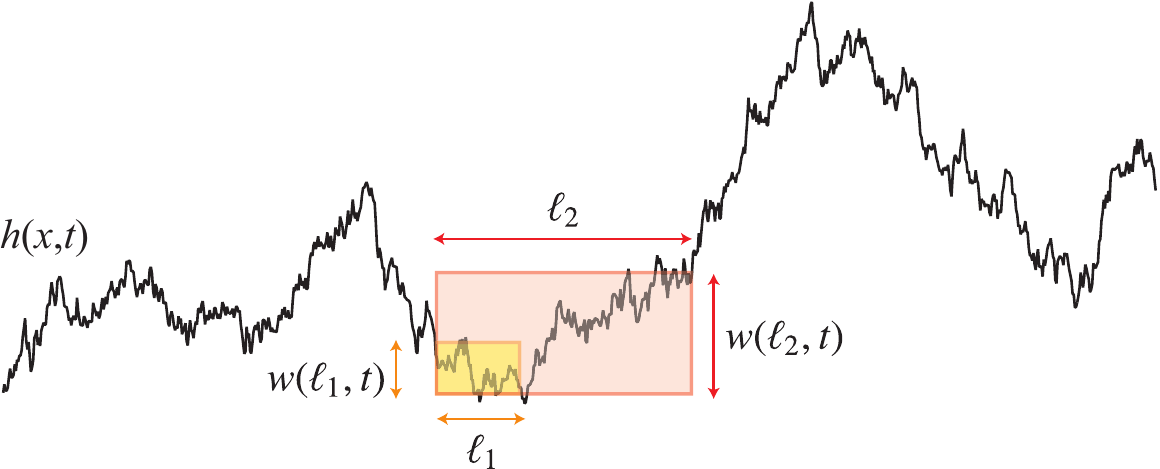}
 \caption{
Sketch of the definition of the interface width $w(\ell,t)$.
The standard deviation of $h(x,t)$ is calculated
 within a window of lateral size $\ell$.
It is then averaged over space (by shifting the window)
 and realizations, and this defines the interface width $w(\ell,t)$.
}
 \label{fig2}
\end{figure}%

In the literature, such kinetic roughening of interfaces has been reported
 in a wide variety of growth processes
 \cite{Barabasi.Stanley-Book1995,HalpinHealy.Zhang-PR1995,Takeuchi-JSM2014},
 classical examples being bacteria colonies on agar, slow combustion of paper,
 growth of solid thin film, etc.
Kinetic roughening seems to be a common feature of interfaces
 that grow through short-ranged interactions under the influence of noise,
 whether it is inherent to dynamics (such as cell division)
 or results from heterogeneity of the system
 (such as fibrous structure of paper)%
\footnote{
In the presence of non-local interactions,
 kinetics and resulting morphology of interfaces can be very different.
For example, if particles can diffuse freely in the surrounding environment
 before they adhere to the interface, the growth is controlled
 by the density field of diffusing particles, which is affected
 by the global structure of the interface.
Fractals tend to appear in such a case;
 see \cite{Barabasi.Stanley-Book1995}.
}.
Those interfaces are not merely similar in the qualitative sense,
 but in many cases they have a striking quantitative property,
 namely, scale invariance.
The example of cancer cell colony growth in \figref{fig1}(a)
 beautifully shows that, as time $t$ is increased,
 both the amplitude of interface fluctuations $\delta h$
 and the length of correlation in the spanwise direction $x$
 seem to grow.
To quantify this, we define the height of interface $h(x,t)$
 as a single-valued function of spanwise position $x$ and time $t$
 (\figref{fig1}(a))%
\footnote{
By using the single-valued function $h(x,t)$,
 we ignore overhangs of the interfaces.
In the literature, a growth equation for an interface of general shape,
 including overhangs, was proposed \cite{Maritan.etal-PRL1992}.
On the basis of numerical results
 presented in that paper \cite{Maritan.etal-PRL1992},
 the effect of overhangs seems to be minor.
A few later numerical studies
 \cite{Marsili.Zhang-PRE1998,RodriguezLaguna.etal-JSM2011}
 also reached the conclusion that overhangs do not seem to change
 the scaling laws of growing interfaces that we are interested in here.
The use of the single-valued height $h(x,t)$ can be justified
 by these observations, and is also favored for the sake of simplicity.
For those who analyze interface data,
 overhangs may simply be removed by taking, e.g.,
 the maximum, the mean value, etc., of all the interface heights
 detected at each position $x$.
Note, however, that there do exist other situations 
 in which overhangs matter, especially if interfaces are pinned
 \cite{Grassberger-a2017}.
Therefore, from the experimental viewpoint, it may be worth trying
 analysis that takes account of overhangs,
 whenever they seem to have large-scale structure.
}
 and measure the standard deviation of $h(x,t)$
 within windows of lateral size $\ell$,
 which is called the interface width $w(\ell,t)$ (\figref{fig2}).
In many experiments, as well as in toy models of growth processes,
 the following power law has been found for this $w(\ell,t)$
 \cite{Barabasi.Stanley-Book1995,HalpinHealy.Zhang-PR1995},
 called the Family-Vicsek scaling%
\footnote{\label{ft:HDiffCorr}
Alternatively, one can also use the height-difference correlation function
 $C_\mathrm{h}(\ell,t) \equiv \expct{[h(x+\ell,t)-h(x,t)]^2}$,
 which scales in the same way as $w(\ell,t)^2$
 (with a different scaling function).
}:
\begin{equation}
 w(\ell,t) \sim t^{\,\beta} F_w(\ell t^{-1/z}) \sim \begin{cases} \ell^\alpha & \text{for $\ell \ll \xi(t)$}, \\ t^{\,\beta} & \text{for $\ell \gg \xi(t)$}, \end{cases} \qquad
 \text{with $\xi(t) \sim t^{1/z} ~(z \equiv \alpha/\beta)$,}
  \label{eq:FamilyVicsek}
\end{equation}
 and a scaling function $F_w(\cdot)$.
This implies that the statistical properties
 of $\delta h(x,t) \equiv h(x,t) - \expct{h(x,t)}$ are kept invariant
 under the following scale transformation:
\begin{equation}
 x \mapsto bx, \quad t \mapsto b^z t, \quad \delta h \mapsto b^\alpha \delta h.
 \label{eq:ScaleTrans}
\end{equation} 
Therefore, the set of the exponents $\alpha, \beta, z$
 characterizes kinetic roughening,
 playing the role equivalent to the critical exponents of phase transitions.
One may then expect that kinetic roughening can also be classified
 in terms of universality classes, which share
 the same values of the scaling exponents $\alpha, \beta, z$.
This is indeed the case, at least theoretically%
\footnote{\label{ft:Experiments}
Unfortunately, universality is not as clear in experimental studies
 \cite{Barabasi.Stanley-Book1995,HalpinHealy.Zhang-PR1995},
 possibly because of some complexities in experimental systems
 and/or often limited length and time scales of observation,
 as well as limited statistical accuracy.
However, recently, a growing number of experiments have shown
 exponent values in agreement with known universality classes,
 in particular the KPZ class \cite{Takeuchi-JSM2014}.
The cancer colony growth \cite{Huergo.etal-PRE2012}
 and the particle deposition \cite{Yunker.etal-PRL2013}
 illustrated in \figref{fig1} are among such experiments.
},
 so that several universality classes and corresponding continuum equations
 have been proposed in the literature \cite{Barabasi.Stanley-Book1995}.

\section{Continuum equations and universality classes}  \label{sec:Continuum}

\subsection{Edwards-Wilkinson equation}

Now we attempt to construct, on the phenomenological basis,
 a continuum equation for the interface height $h(\vx,t)$
 growing on a $d$-dimensional substrate, i.e., $\vx \in \mathbb{R}^d$
 and $h, t \in \mathbb{R}$.
The symmetry we expect for kinetic roughening is
 (i) invariance under time translation $t \mapsto t + t_0$,
 (ii) invariance under space translation $\vx \mapsto \vx + \vx_0$,
 (iii) invariance under space inversion and rotation, e.g., 
 $\vx \mapsto -\vx$, and
 (iv) invariance under height translation $h \mapsto h + h_0$.
The possibly simplest such equation
 is the following linear partial differential equation,
 called the Edwards-Wilkinson (EW) equation:
\begin{equation}
 \prt{}{t}h(\vx,t) = v_0 + \nu \vnabla^2 h + \eta(\vx,t).  \label{eq:EWeq}
\end{equation}
Here, $\eta(\vx,t)$ is white Gaussian noise such that
\begin{equation}
 \expct{\eta(\vx,t)} = 0, \qquad
 \expct{\eta(\vx,t)\eta(\vx',t')} = D \delta(\vx-\vx')\delta(t-t')
 \label{eq:Noise}
\end{equation}
 with ensemble average $\expct{\cdots}$,
 and $v_0, \nu, D$ are constant parameters.
The second term of \eqref{eq:EWeq} corresponds
 to surface tension, or diffusion, which tends to smooth
 irregularities of the interface.
The first term is simply a constant driving force.
Since this term can be trivially removed by Galilean transformation
 $h \mapsto h + v_0 t$, i.e., by using the comoving frame,
 it is often omitted in the theoretical literature.
As a result, the EW equation has an additional height inversion symmetry,
 i.e., (v) invariance under $h \mapsto -h$,
 in addition to the symmetries (i)-(iv) listed above.

To obtain the scaling exponents $\alpha, \beta, z$
 for the EW equation, we may impose the invariance
 under the scale transformation \pref{eq:ScaleTrans}
 directly upon \eqref{eq:EWeq} with $h(x,t) = v_0t + \delta h(x,t)$.
This leads to $b^{\alpha-z} = b^{\alpha-2} = b^{-(d+z)/2}$, hence
\begin{equation}
 \alpha = \frac{2-d}{2}, \quad
 \beta = \frac{2-d}{4}, \quad
 z = 2. \qquad
 \text{(EW class)}  \label{eq:EWexp}
\end{equation}
Although such naive power counting does not always work in general,
 for the EW equation it gives the correct exponent values,
 as one can easily check by solving \eqref{eq:EWeq} in Fourier space
 \cite{Barabasi.Stanley-Book1995,HalpinHealy.Zhang-PR1995}.

Instead of directly solving for the height function $h(x,t)$,
 one can also deal with the probability density
 of the height profile $h(\vx)$ at time $t$, $P[h(\vx),t]$,
 via the Fokker-Planck equation \cite{Gardiner-Book2004}.
For the EW equation \pref{eq:EWeq} with $v_0 = 0$, it reads%
\footnote{
The functional version of the Fokker-Planck equation can be derived
 as follows.
Consider the Langevin equation for multiple particles $X_i(t)$,
\begin{equation*}
 \prt{X_i}{t} = F_i[\{X_j\}] + \eta_i(t),
\end{equation*}
 where $\{X_j\}$ denotes the all-particle configuration 
 and $\eta_i(t)$ is white Gaussian noise with $\expct{\eta_i(t)} = 0$ and
 $\expct{\eta_i(t) \eta_j(t')} = D\delta_{ij}\delta(t-t')$.
The corresponding Fokker-Planck equation is \cite{Gardiner-Book2004}
\begin{equation*}
 \prt{}{t}P[\{X_j\},t] = -\sum_i \prt{}{X_i}F_i[\{X_j\}]P[\{X_j\},t] + \frac{D}{2}\sum_i \prts{}{X_i}{2} P[\{X_j\},t].
\end{equation*}
Now, if we define $X_i \equiv h(x_i)$ with $x_i = i \Delta x$
 and take the continuum limit $\Delta x \to 0$,
 we obtain the desired equation.
}
\begin{equation}
 \prt{}{t} P[h(\vx),t] = -\int \rd^d \vx \varder{}{h} (\nu\vnabla^2 h) P[h(\vx),t] + \frac{D}{2}\int \rd^d\vx \varders{}{h}{2} P[h(\vx),t]  \label{eq:EWFP}
\end{equation}
 with the functional derivative $\varder{}{h}$,
 which formally corresponds to
 $\varder{}{h} = \prt{}{h} - \vnabla \cdot \prt{}{(\vnabla h)}$.
Then the stationary solution to \eqref{eq:EWFP}, $P_\mathrm{stat}[h(\vx)]$,
 can be readily found to be
\begin{equation}
 P_\mathrm{stat}[h(\vx)] \propto \exp\[-\int\rd^d\vx \frac{\nu}{D}(\vnabla h)^2\].  \label{eq:EWFPstat}
\end{equation}
In particular, for $d=1$, it is \textit{exactly} the probability density
 of the Brownian motion with diffusion coefficient $D/4\nu$,
 if $x$ and $h$ for the interface are interpreted as time and position,
 respectively, of the Brownian motion.
In other words, if $B(t)$ denotes the Brownian motion
 with mean squared displacement $\expct{[B(t)-B(0)]^2} = t$
 (Wiener process \cite{Gardiner-Book2004}),
\begin{equation}
 h_\mathrm{stat, 1D}(x) = \sqrt{A} B(x) \qquad (A \equiv D/2\nu)
  \label{eq:BrownianInitCond}
\end{equation}
 gives the statistically stationary solution of the 1D EW equation.
Since
 $\expct{[h_\mathrm{stat, 1D}(x+\ell)-h_\mathrm{stat, 1D}(x)]^2} \propto \ell$,
 this directly accounts for $\alpha = 1/2$ of the 1D EW class
 (see footnote~\ref{ft:HDiffCorr}).

\subsection{Kardar-Parisi-Zhang equation}

Since the EW equation is linear,
 many of its fluctuation properties can be solved exactly.
But for this very reason, the EW equation is not sufficiently generic;
 growth processes in nature are in fact usually nonlinear.
A simple geometric argument is enough to illustrate why growth is nonlinear
 in many cases.

\begin{figure}[t]
 \centering
 \includegraphics[scale=0.8,clip]{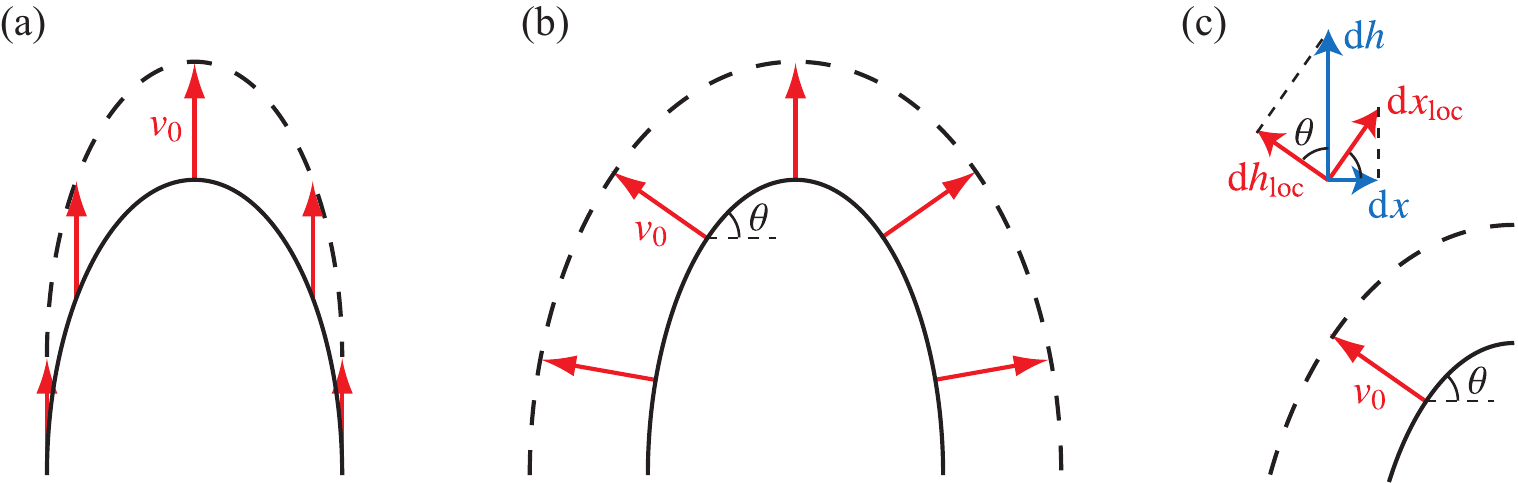}
 \caption{
Sketch of the interface that grows according to the EW equation (a)
 and that grows in the locally normal direction (b).
The effect of noise and diffusion is neglected in this sketch.
The panel (c) illustrates how to transform the local coordinates
 $(h_\mathrm{loc}, x_\mathrm{loc})$ to the global ones $(h,x)$.
This sketch is adapted from Fig.~3.1 in \cite{HalpinHealy.Zhang-PR1995}.
}
 \label{fig3}
\end{figure}%

Let's start from the EW equation \pref{eq:EWeq}.
When we wrote it down, we had fixed the axes and implicitly assumed
 that the interface would always grow along the $h$-axis.
This implies that, even a protrusion of the interface (\figref{fig3}(a))
 would attempt to advance upward at speed $v_0$,
 even though it will also be deformed
 by noise and diffusion of the EW equation.
However, what happens more naturally is that the interface grows
 isotropically, so that each local piece of the interface advances
 in the direction normal to the interface (\figref{fig3}(b)).
To take this effect into account, we may assume that
 the interface is evolved locally by the EW equation,
 with the height and space coordinates $(h_\mathrm{loc}, x_\mathrm{loc})$
 taken locally along the normal and tangential axes, respectively.
Then we transform $(h_\mathrm{loc}, x_\mathrm{loc})$ to the global coordinates
 $(h,x)$ and obtain a new equation.
In order for the new coordinates to describe the desired interface evolution,
 their infinitesimal changes $\rd h$ and $\rd x$ should be
 associated with $\rd h_\mathrm{loc}$ and $\rd x_\mathrm{loc}$
 in the following manner (\figref{fig3}(c)):
\begin{equation}
 \rd h
 = \rd h_\mathrm{loc} / \cos\theta
 = \rd h_\mathrm{loc} \sqrt{1+(\vnabla h)^2}, \qquad
 \rd x
 = \rd x_\mathrm{loc} \cos\theta
 = \rd x_\mathrm{loc} / \sqrt{1+(\vnabla h)^2},
\end{equation}
 where $\theta$ is the local inclination angle,
 hence $\tan\theta = |\vnabla h|$.
By inserting it to the evolution of $(h_\mathrm{loc}, x_\mathrm{loc})$,
 i.e., the EW equation \pref{eq:EWeq}, and keeping only the lowest-order
 nonlinearity for small $|\vnabla h|$, we obtain
\begin{equation}
 \prt{}{t}h(\vx,t) = v_0 + \nu \vnabla^2 h + \frac{\lambda}{2} (\vnabla h)^2 + \eta(\vx,t),  \label{eq:KPZeq}
\end{equation}
 where the noise $\eta(\vx,t)$ is still given by \eqref{eq:Noise}.
This is the KPZ equation, the (main) protagonist of these lecture notes.
Note that the effect of the local slope discussed here
 is only one possible source of nonlinearity;
 in general there can be others, such as nonlinearity in interactions
 between constituent particles.
Most probably, they also contribute to the $(\vnabla h)^2$ term,
 which is indeed the lowest-order nonlinearity
 allowed by the symmetry requirements (i)-(iv)
 discussed in the previous section.
Therefore, the KPZ equation is not necessarily a model of isotropic interfaces,
 but is a more generic model of interfaces
 growing under some nonlinear mechanisms.
The coefficient $\lambda$ of the $(\vnabla h)^2$ term
 should be regarded as an independent parameter of the equation.

Some remarks are now in order.
First, the constant term $v_0$ can still be got rid of
 by the Galilean transformation,
 so that we can set $v_0 = 0$ without loss of generality.
Second, the nonlinear term $(\vnabla h)^2$
 breaks the height inversion symmetry (v) that the EW equation possesses.
In this sense, the KPZ equation describes a \textit{bona fide} growth process,
 which is not simply due to the constant term $v_0$ that can be removed.
Finally, the newly added nonlinear term is relevant
 in the renormalization group sense
 \cite{Kardar.etal-PRL1986,Barabasi.Stanley-Book1995,HalpinHealy.Zhang-PR1995},
 at least for the physically relevant dimensionalities $d \leq 2$.
The KPZ equation therefore represents a new universality class, the KPZ class.
For 1D, the values of the scaling exponents can be determined exactly to be
\begin{equation}
 \alpha = 1/2, \quad \beta = 1/3, \quad z = 3/2 \qquad
 (\text{1D KPZ class})  \label{eq:1DKPZexp}
\end{equation}
 because of certain symmetries
 that will be described in Section \ref{sec:KPZeq}.
In contrast, no exact values are known for higher dimensions;
 for $d=2$, a recent numerical estimation \cite{Pagnani.Parisi-PRE2015}
 gave $\alpha = 0.3869(4)$, where the number in the parentheses indicates
 the uncertainty, from which $\beta = 0.2398(3)$ and $z = 1.6131(4)$ follow
 (using \eqref{eq:KPZScalingRelation} explained below)%
\footnote{
Note however that estimates from different studies still vary a little.
See Table I in \cite{Pagnani.Parisi-PRE2015}
 as well as more recent work \cite{Kelling.etal-a2017}.
}.
Note that the naive power counting we did for the EW equation
 does not work here, because in general coefficients such as $\nu, \lambda, D$
 can also change by the scale transformation \pref{eq:ScaleTrans}.
For the KPZ equation, dynamic renormalization group calculation
 \cite{Kardar.etal-PRL1986,Barabasi.Stanley-Book1995,HalpinHealy.Zhang-PR1995}
 indeed shows that $\nu$ and $D$ change with scale,
 while $\lambda$ is actually scale-invariant.
We will come back to this point in Section \ref{sec:KPZ-Burgers}.

\subsection{Mullins-Herring equation and its nonlinear variant}

In the literature, there have been proposed
 a few other continuum equations and associated universality classes
 for kinetic roughening of interfaces.
For example, if deposited particles have enough time to diffuse
 on the surface until the next particle is deposited,
 one obtains the following Mullins-Herring equation
 within the linear theory \cite{Barabasi.Stanley-Book1995}:
\begin{equation}
 \prt{}{t}h(\vx,t) = v_0 - K \vnabla^4 h + \eta(\vx,t).  \label{eq:MHeq}
\end{equation}
The nonlinear variant of this model reads
\begin{equation}
 \prt{}{t}h(\vx,t) = v_0 - K \vnabla^4 h + \lambda_1 \vnabla^2 (\vnabla h)^2 + \lambda_2 \vnabla \cdot (\vnabla h)^3 + \eta(\vx,t),  \label{eq:MBEeq}
\end{equation}
 where the second nonlinear term $\lambda_2 \vnabla \cdot (\vnabla h)^3$
 is often neglected,
 because its physical origin is unclear \cite{Barabasi.Stanley-Book1995}.
The corresponding universality classes are sometimes called
 the linear (\eqref{eq:MHeq}) and nonlinear (\eqref{eq:MBEeq})
 molecular beam epitaxy classes.
The interested readers are referred to other reviews and textbooks,
 such as \cite{Barabasi.Stanley-Book1995}.

\subsection{Growth under quenched noise and the quenched KPZ equation}

For some experimental examples of growing interfaces,
 such as wetting fronts of paper \cite{Barabasi.Stanley-Book1995},
 noise comes from heterogeneity of the medium.
Then it may be more appropriate to replace the spatiotemporal noise
 $\eta(\vx,t)$ that we have used in continuum equations
 by quenched noise $\eta_\mathrm{q}(\vx,h)$, which is now
 a function of $\vx$ and $h$ and does not depend explicitly on time.
If we do so for the KPZ equation, we obtain the quenched KPZ (qKPZ) equation:
\begin{equation}
 \prt{}{t}h(\vx,t) = F + \nu \vnabla^2 h + \frac{\lambda}{2} (\vnabla h)^2 + \eta_\mathrm{q}(\vx,h(\vx,t)).  \label{eq:qKPZeq}
\end{equation}
 with
\begin{equation}
 \expct{\eta_\mathrm{q}(\vx,h)\eta_\mathrm{q}(\vx',h')} = \delta(\vx-\vx')\Delta_\kappa(h-h'),
\end{equation}
 where $\Delta_\kappa(r)$ is an even function of $r$
 that monotonically decreases to zero over a finite distance $\kappa$
 (the delta correlation corresponds to $\kappa \to 0$)%
\footnote{
The delta correlation limit $\kappa\to 0$ is not taken here,
 because the critical force $F_\mathrm{c}$
 of the pinning-depinning transition is known to diverge in this limit
 \cite{Barabasi.Stanley-Book1995}.
}.
The constant term of \eqref{eq:qKPZeq} is denoted by $F$ (for ``force'')
 following the convention,
 which \textit{cannot} be got rid of by Galilean transformation
 because of the explicit dependence of noise on $h$.

An important characteristic of the qKPZ equation is that
 it shows a pinning-depinning transition as the force $F$ is varied.
More specifically, if $F$ is large enough,
 the interface will grow unceasingly.
In this case $\eta_\mathrm{q}(\vx,h(\vx,t))$ is equivalent to $\eta(\vx,t)$,
 so that the interface is described by the KPZ scaling laws.
In contrast, if $F$ is small, there will be fractions of the interface
 that experience too strong negative noise $\eta_\mathrm{q}(\vx,h(\vx,t))$
 to be overcome by the given driving force $F$.
Once they stop, they may remain pinned,
 because the value of $\eta_\mathrm{q}(\vx,h(\vx,t))$ does not change
 unless the interface moves.
In this way, the interface will be entirely pinned eventually.
Then, those pinning and depinning phases are separated
 by a critical force $F_\mathrm{c}$.
At $F = F_\mathrm{c}$ the interface
 is governed by a distinct set of scaling laws.
It is also seen if $F$ is not exactly $F_\mathrm{c}$,
 up to a crossover time, which can be longer
 than the finite observation time of experiments or simulations.
Therefore, in practice, the scaling laws of critically pinned interfaces
 may appear in a region of parameter space.

In the literature, critically pinned interfaces
 have been studied with various models.
The situation is rather involved and not fully understood;
 a few universality classes have been argued to arise in different situations
 (see \cite{Barabasi.Stanley-Book1995,Grassberger-a2017}
 and references therein).
It seems to be important (but not only)
 whether or not overhangs are allowed and relevant,
 and whether or not nonlinearity is relevant at the critical point.
Concerning the latter point, even among models with nonlinearity
 [corresponding to $\frac{\lambda}{2} (\vnabla h)^2$ in \eqref{eq:qKPZeq}],
 for some models nonlinearity is relevant at the critical point,
 and for others it is not.
More specifically, the coefficient $\lambda$ was argued to behave
 as $\lambda \sim (F-F_\mathrm{c})^{-\phi}$ \cite{Barabasi.Stanley-Book1995},
 so that it may diverge ($\phi>0$) or vanish ($\phi<0$) at the critical point.
The directed percolation depinning model \cite{Barabasi.Stanley-Book1995}
 is an example where overhangs are forbidden and nonlinearity is relevant,
 and is considered to represent the qKPZ universality class.
In particular, for 1D, this model makes a connection to the critical behavior
 of the directed percolation universality class
 for absorbing-state phase transitions \cite{Hinrichsen-AP2000}.
As a result, the scaling exponents of the 1D qKPZ class are given by
 $\alpha = \beta = \nu_\perp^\mathrm{DP} / \nu_\parallel^\mathrm{DP} \approx 0.633$ and $z=1$ \cite{Barabasi.Stanley-Book1995},
 where $\nu_\perp^\mathrm{DP}$ and $\nu_\parallel^\mathrm{DP}$
 are the critical exponents for diverging correlation length and time
 of the directed percolation class, respectively.
In contrast, if we set $\lambda = 0$ in \eqref{eq:qKPZeq},
 it is called the quenched Edwards-Wilkinson (qEW) equation,
 and the corresponding class is the qEW class.
It is believed to describe the case
 in which both overhangs and nonlinearity are absent or irrelevant.
If overhangs are allowed and relevant, the critical interface may become
 fractal and described by the ordinary isotropic percolation class
 in $(d+1)$ spatial dimensions
 \cite{Ji.Robbins-PRB1992,Bizhani.etal-PRE2012,Grassberger-a2017}.
Some models, in particular the random field Ising model
 without spontaneous nucleation
 \cite{Ji.Robbins-PRB1992,Barabasi.Stanley-Book1995},
 show a transition between the fractal and self-affine morphology
 (akin to qEW) for $d \geq 2$ (i.e., total spatial dimensionality $\geq 3$)
 \cite{Ji.Robbins-PRB1992,Bizhani.etal-PRE2012,Barabasi.Stanley-Book1995}.
This transition is characterized by the tricritical percolation
 \cite{Janssen.etal-PRE2004,Bizhani.etal-PRE2012}.

\section{Basic properties of the KPZ equation}  \label{sec:KPZeq}

In this section we concentrate on the KPZ equation \pref{eq:KPZeq}
 and study its basic properties.
In the following, we set $v_0 = 0$ without loss of generality.

\subsection{Relation to the noisy Burgers equation}  \label{sec:KPZ-Burgers}

Already since the invention of the KPZ equation \cite{Kardar.etal-PRL1986},
 its relation to a stirred fluid problem has been recognized%
\footnote{
In fact, the basic scaling properties of the KPZ equation were studied
 before it was invented,
 in the context of the noisy Burgers equation \cite{Forster.etal-PRA1977}.
}.
To see this, take the gradient of the KPZ equation \pref{eq:KPZeq}
 and define $\vv(\vx,t) \equiv -\lambda\vnabla h(\vx,t)$, then we obtain
\begin{equation}
 \prt{\vv}{t} + (\vv \cdot \vnabla)\vv = \nu \vnabla^2 \vv - \lambda \vnabla \eta(\vx,t).  \label{eq:noiseBurgers}
\end{equation}
It is the noisy Burgers equation,
 known as a simple theoretical model of randomly stirred fluid.
As such, it satisfies a symmetry expected for fluid,
 namely the invariance under Galilean transformation,
 $\vv(\vx,t) \mapsto \vv_0 + \vv(\vx-\vv_0 t, t)$
 with a constant vector $\vv_0$, for which we also use
 $\expct{\eta(\vx,t)\eta(\vx',t')} = \expct{\eta(\vx-\vv_0 t,t)\eta(\vx'-\vv_0 t',t')} = D\delta(\vx-\vx')\delta(t-t')$.
In terms of the interface height $h(\vx,t)$,
 this translates into the invariance under the following tilt transformation
 with any constant vector $\bm{s}$:
\begin{equation}
 h_\mathrm{new}(\vx_\mathrm{new},t) = h(\vx,t) + \bm{s} \cdot \vx - \frac{\lambda}{2} \bm{s}^2 t, \qquad
 \vx_\mathrm{new} = \vx - \lambda \bm{s}t,  \label{eq:STS}
\end{equation}
 where the same averaging over noise is used.
This is called the statistical tilt symmetry of the KPZ equation.
Moreover, since the Galilean symmetry
 should remain unchanged under the scale transformation \pref{eq:ScaleTrans},
 we can conclude that the coefficient $\lambda$ should be scale-invariant%
\footnote{
To see this clearly, following \cite{Forster.etal-PRA1977},
 one may introduce a coefficient $\lambda_0 = 1$
 to the advection term $(\vv \cdot \vnabla)\vv$.
Then the system is invariant under
 $\vv(\vx,t) \mapsto \vv_0 + \vv(\vx-\lambda_0\vv_0 t, t)$,
 which physically corresponds to the Galilean transformation
 only if $\lambda_0 = 1$.
This implies that $\lambda_0$ remains unity
 even after the scale transformation \pref{eq:ScaleTrans};
 in other words, $\lambda_0$ is scale-invariant.
Now, if we define the velocity field anew
 by $\vv'(\vx,t) \equiv -\vnabla h(\vx,t)$,
 the noisy Burgers equation reads
 $\prt{\vv'}{t} + \lambda (\vv' \cdot \vnabla)\vv' = \nu \vnabla^2 \vv' - \vnabla \eta(\vx,t)$.
As we have just learned, the coefficient of the advection term,
 which is now $\lambda$, is scale-invariant.
}.
By applying the scale transformation \pref{eq:ScaleTrans}
 to the KPZ equation \pref{eq:KPZeq}, this scale invariance of $\lambda$
 results in the following scaling relation,
\begin{equation}
 \alpha + z = 2,  \label{eq:KPZScalingRelation}
\end{equation}
 which is valid for the KPZ class in any dimension $d$.

\subsection{Stationary probability density for the 1D KPZ equation and fluctuation-dissipation theorem}

As we did for the EW equation, we can also write down
 the Fokker-Planck equation for the KPZ equation \pref{eq:KPZeq}.
This reads:
\begin{equation}
 \prt{}{t} P[h(\vx),t] = -\int \rd^d \vx \varder{}{h} \[ \nu\vnabla^2 h + \frac{\lambda}{2}(\vnabla h)^2 \] P[h(\vx),t] + \frac{D}{2}\int \rd^d\vx \varders{}{h}{2} P[h(\vx),t].  \label{eq:KPZFP}
\end{equation}
Interestingly, for 1D (only), one can easily check that
 the stationary solution \pref{eq:EWFPstat} for the EW equation,
 i.e., the probability density of the Brownian motion,
 is also the stationary solution for the KPZ equation%
\footnote{
This is because
 $\int\rd x \varder{}{h} (\nabla h)^2 P_\mathrm{stat}[h(x),t] = 0$
 for 1D (but not otherwise).
}.
This immediately gives $\alpha = 1/2$.
Using the scaling relation \pref{eq:KPZScalingRelation}
 and $z \equiv \alpha/\beta$, we obtain $\beta = 1/3$ and $z = 3/2$,
 the complete set of the scaling exponents of the 1D KPZ class.
The fact that the Brownian motion gives the stationary probability density
 of the 1D KPZ equation is actually a consequence
 of the fluctuation-dissipation theorem \cite{Deker.Haake-PRA1975}
 that it satisfies.
This implies that the field theory underlying the 1D KPZ equation
 is much simpler than its higher-dimensional counterparts
 \cite{Deker.Haake-PRA1975}.

\subsection{Well-definedness of the KPZ equation and relation to the stochastic heat equation}

After all the explanations given on the KPZ equation, 
 it is, in fact, known to be ill-defined
 \cite{Bertini.Giacomin-CMP1997,Sasamoto.Spohn-PRL2010,Sasamoto-PTEP2016},
 even if we should decide not to care about mathematical subtleties.
It is easy to see from the form of the KPZ equation \pref{eq:KPZeq},
 because, on the one side it has delta-correlated noise $\eta(\vx,t)$,
 which is non-differentiable, and on the other side
 it asks to compute the squared gradient, $(\vnabla h)^2$,
 which would diverge if the operation were carried out literally.

To circumvent this problem,
 we resort to the mapping to another stochastic process.
Using the Cole-Hopf transformation
\begin{equation}
 Z(\vx,t) \equiv \exp\[\frac{\lambda}{2\nu}h(\vx,t)\],
  \label{eq:ColeHopfTrans}
\end{equation}
 we can \textit{formally} rewrite the KPZ equation \pref{eq:KPZeq}
 to the following, stochastic heat equation:
\begin{equation}
 \prt{}{t} Z(\vx,t) = \nu \vnabla^2 Z(\vx,t) + \frac{\lambda}{2\nu}Z(\vx,t) \text{~``$\times$''~} \eta(\vx,t),  \label{eq:SHE1}
\end{equation}
 where the meaning of ``$\times$'' will be explained soon.
This equation can also be expressed as $\rd Z(\vx,t) = \nu \vnabla^2 Z(\vx,t) \rd t + \frac{\lambda}{2\nu}Z(\vx,t) \text{~``$\times$''~} \rd B_\eta(\vx,t)$
 with $\rd B_\eta(\vx,t) \equiv \eta(\vx,t) \rd t$.
Compared with the KPZ equation, the stochastic heat equation may look simpler
 because it is linear, but it is deceptively so,
 as we have the multiplicative noise term instead.
According to stochastic calculus \cite{Gardiner-Book2004}, whenever we have
 such multiplicative noise in a stochastic differential equation,
 we need to specify the definition of the product
 $Z(\vx,t) \text{~``$\times$''~} \rd B_\eta(\vx,t)$.
If we choose the Stratonovich product%
\footnote{
For the stochastic differential equation
 $\rd Z(t) = F(Z(t),t) \rd t + G(Z(t),t) \text{~``$\times$''~} \rd B(t)$
 with infinitesimal increment $\rd B(t) \equiv B(t + \rd t) - B(t)$
 of the Wiener process $B(t)$,
 the Stratonovich product $G(Z(t),t) \circ \rd B(t)$
 is defined as the continuum limit
 of $G(\tfrac{Z(t_{i+1}) + Z(t_i)}{2}, t_i)[B(t_{i+1}) - B(t_i)]$,
 and the It\^o product $G(Z(t),t) \rd B(t)$ is the continuum limit
 of $G(Z(t_i), t_i)[B(t_{i+1}) - B(t_i)]$ \cite{Gardiner-Book2004}.
While for the Stratonovich product we can use the usual chain rule
 of differentiation, the It\^o product has a great advantage that
 the two operands $G(Z(t_i), t_i)$ and $B(t_{i+1}) - B(t_i)$
 are statistically independent.
} $Z(\vx,t) \circ \rd B_\eta(\vx,t)$,
 the usual chain rule of differentiation can be used,
 so that the stochastic heat equation \pref{eq:SHE1} is equivalent
 to the KPZ equation \pref{eq:KPZeq}, but then both of them are ill-defined.

To see this clearly \cite{Sasamoto-PTEP2016,Sasamoto.Spohn-PRL2010},
 let's consider the 1D case
 and replace the noise $\eta(x,t)$ by colored Gaussian noise $\eta_\kappa(x,t)$
 such that
\begin{equation}
 \expct{\eta_\kappa(x,t)} = 0, \qquad
 \expct{\eta_\kappa(x,t)\eta_\kappa(x',t')} = D\Delta_\kappa(x-x')\delta(t-t'),
  \label{eq:SmoothNoise1}
\end{equation}
 with
\begin{equation}
 \Delta_\kappa(x) \equiv \frac{1}{\kappa\sqrt{\pi}}\e^{-(x/\kappa)^2} \to \delta(x) \qquad (\kappa \to 0).  \label{eq:SmoothNoise2}
\end{equation}
Using the conversion formula between
 the Stratonovich and It\^o products \cite{Gardiner-Book2004}, we have
\begin{equation}
 Z(x,t) \circ \rd B_{\eta_\kappa}(x,t) = Z(x,t) \rd B_{\eta_\kappa}(x,t) + \frac{1}{2}\rd Z(x,t) \rd B_{\eta_\kappa}(x,t),  \label{eq:Ito2Stratonovich}
\end{equation}
 where the products in the right hand side are now It\^o's ones.
Then we obtain
\begin{equation}
 \expct{\rd Z(x,t) \rd B_{\eta_\kappa}(x,t)}
 = \frac{\lambda}{2\nu} \expct{Z(x,t)} \expct{\rd B_{\eta_\kappa}(x,t)^2}
 = \frac{\lambda}{2\nu} \expct{Z(x,t)} D\Delta_\kappa(0)\rd t
 = \frac{\lambda}{2\nu} \expct{Z(x,t)} \frac{D}{\kappa\sqrt{\pi}} \rd t,
  \label{eq:StratonovichDivergence}
\end{equation}
 which diverges in the limit $\kappa \to 0$.
On the other hand, this singularity is not too pathologic
 because, if $\expct{Z(x,t)}$ is uniform in space,
 it simply amounts to uniform translation in $h$.
It can therefore be removed by riding on the comoving frame,
 or, equivalently, using the It\^o product from the beginning.
Indeed, the stochastic heat equation with the It\^o product
\begin{equation}
 \prt{}{t} Z(x,t) = \nu \vnabla^2 Z(x,t) + \frac{\lambda}{2\nu}Z(x,t) \eta(x,t),  \label{eq:SHE}
\end{equation}
 can be shown to be mathematically well-defined, at least for 1D \cite{Bertini.Giacomin-CMP1997,Sasamoto.Spohn-PRL2010,Sasamoto.Spohn-NPB2010,Amir.etal-CPAM2011,Sasamoto-PTEP2016}.
Therefore, we can start from \eqref{eq:SHE} and \textit{define}
\begin{equation}
 h(x,t) \equiv \frac{2\nu}{\lambda}\log Z(x,t).  \label{eq:KPZeqHeightDef}
\end{equation}
Suppose a solution $Z(x,t)$ to \eqref{eq:SHE} is obtained,
 $h(x,t)$ defined by \eqref{eq:KPZeqHeightDef} is formally called
 the ``solution of the KPZ equation'' in the literature%
\footnote{
As an alternative approach, Hairer extended what is called
 the rough path theory in mathematics and successfully provided
 a mathematical foundation to the KPZ equation \cite{Hairer-AM2013}.
This approach was then further developed to the theory of regularity structures
 \cite{Hairer-IM2014}.
For this series of work,
 Hairer was awarded a Fields Medal of the year 2014.
}.

\section{Exact results on the 1D KPZ class 1: polynuclear growth model}
\label{sec:Exact1}

From the beginning of its history,
 the KPZ class has been extensively studied
 as an instance of non-equilibrium scaling laws.
However, it is since the year 2000 that our understanding of the KPZ class
 has been dramatically changed, when a few models in the 1D KPZ class
 were solved with mathematical rigor for the first time
 \cite{Johansson-CMP2000,Prahofer.Spohn-PRL2000}.
The first results were obtained on the one-point distribution
 of the height $h(x,t)$.
The developments have soon extended to other statistical quantities,
 such as the correlation function in space,
 as well as to more generalized models in the 1D KPZ class.
This led to further breakthroughs, in particular
 the establishment of exact solutions to the 1D KPZ equation proper since 2010
 \cite{Sasamoto.Spohn-PRL2010,Sasamoto.Spohn-NPB2010,Amir.etal-CPAM2011,Calabrese.etal-EL2010,Dotsenko-EL2010,Calabrese.LeDoussal-PRL2011,Imamura.Sasamoto-PRL2012,Borodin.etal-MPAG2015}. 
These activities from mathematics and mathematical physics
 continue marking rapid progress \cite{Kriecherbauer.Krug-JPA2010,Corwin-RMTA2012,HalpinHealy.Takeuchi-JSP2015,Quastel.Spohn-JSP2015,Sasamoto-PTEP2016}.
While they are certainly interesting and important in their own,
 they have also conveyed a lot of physical implications,
 which non-specialists may overlook
 when reading mathematical literature.
The aim of the present and following sections is, therefore, to illustrate
 some important rigorous results with emphasis put
 on physical implications and intuitions,
 which allow the author to work in this field
 by experimental and numerical approaches.

In the present section,
 we concentrate on a model called the polynuclear growth (PNG) model,
 which is one of the first models that were solved exactly.
By following Pr\"ahofer and Spohn's celebrated paper
 \cite{Prahofer.Spohn-PRL2000}, it will be illustrated how the model is related
 to a number of different problems, how the exact results were derived,
 and what the outcome is from the physics viewpoint.
In particular, it will turn out that the KPZ class splits into at least
 a few \textit{universality subclasses} as the author calls,
 for which the scaling exponents are the same (\eqref{eq:1DKPZexp})
 but nevertheless other statistical properties of the height fluctuations,
 in particular their distribution and correlation properties,
 are different.

\subsection{Definition of the 1D PNG model}

\begin{figure}[t]
 \centering
 \includegraphics[scale=0.8,clip]{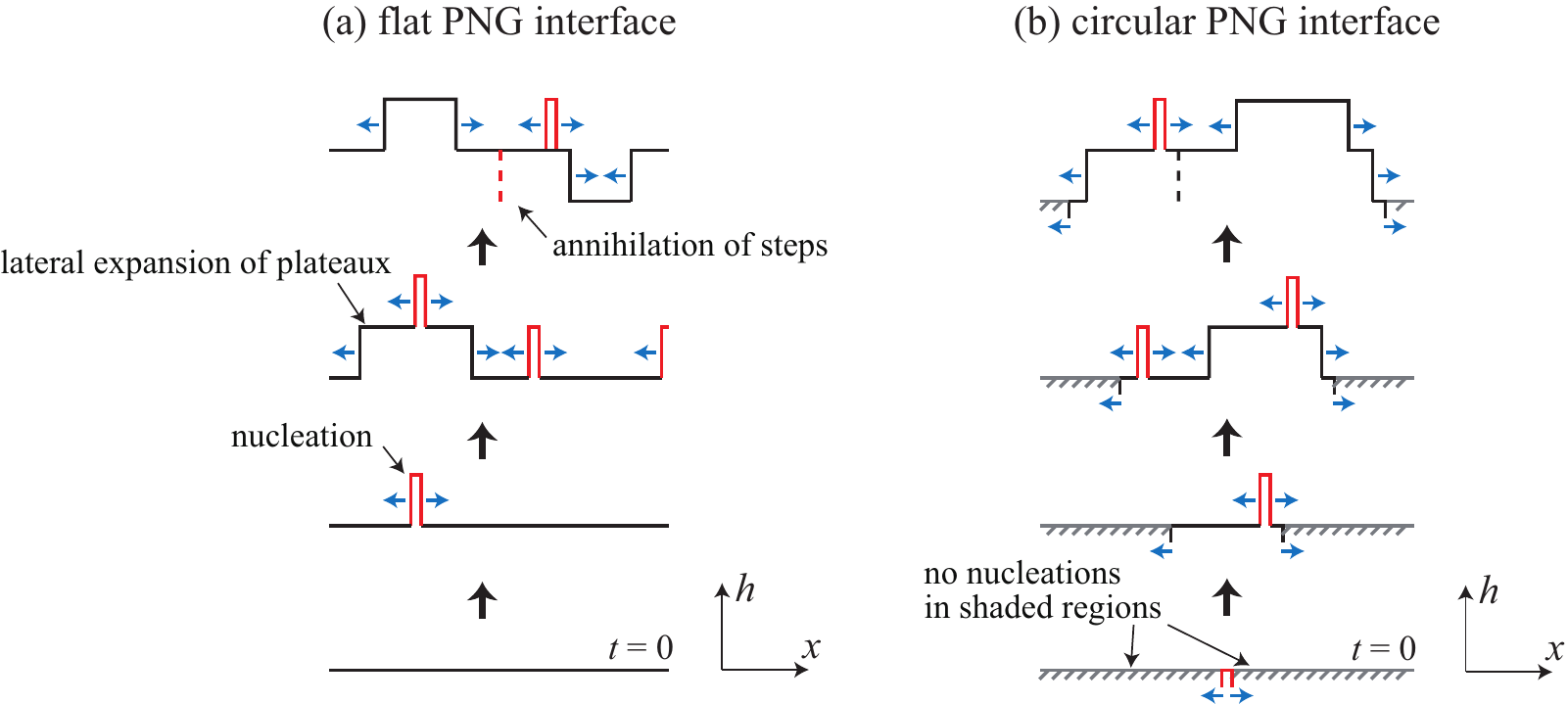}
 \caption{
Sketch of a flat (a) and a circular (b) PNG interface.
In the circular case (b), nucleations do not occur in the shaded regions,
 $|x| > t$.
}
 \label{fig4}
\end{figure}%

The PNG model is a discrete model in the KPZ class,
 which describes a growth process driven by random creations of nuclei
 (\figref{fig4}(a)).
For 1D, $x,t \in \mathbb{R}$ and $h(x,t) \in \mathbb{Z}$.
Starting from $h(x,0)=0$ (bottom sketch of \figref{fig4}(a)),
 there is a probability $\rho$ over unit space and unit time
 that a point nucleus is generated (second sketch).
If this happens at position $x_n$ and time $t_n$,
 the height $h(x_n,t_n)$ is incremented by one, i.e.,
 $h(x_n,t_n) \mapsto h(x_n,t_n)+1$.
Then the steps at both ends of the nucleus separate
 laterally, each at constant speed $\varv$ (third sketch).
When two steps from different nuclei meet each other,
 they simply annihilate (top sketch).
These ingredients (except the random nucleation) can be expressed
 by a single equation,
 $h(x, t+\rd t) = \max_{x-\varv\rd t \leq x' \leq x+\varv\rd t} h(x',t)$.
Nucleations can also occur on top of the expanding plateaux.
As a result, the PNG model, without any constraint, produces
 an on average flat interface that grows upward.
In the following, we set $\rho = 1$ and $\varv=1$ without loss of generality.
Note that we consider an infinitely large system,
 so that we do not need to set the boundary condition.

\subsection{PNG circular interface}  \label{sec:PNGcirc}

While the flat case depicted in \figref{fig4}(a) is simplest
 in view of the definition of the model,
 its variant for curved interfaces actually corresponds
 to the most fundamental case for our purpose.
To deal with a curved interface in the PNG model,
 one can impose a constraint that nucleations can occur
 only within the region $|x| \leq t$ (see \figref{fig4}(b)).
Then the generated interface is bounded by the two extremities
 that are fixed at $h(\pm t, t)=0$, hence it is curved.

\begin{figure}[t]
 \centering
 \includegraphics[clip]{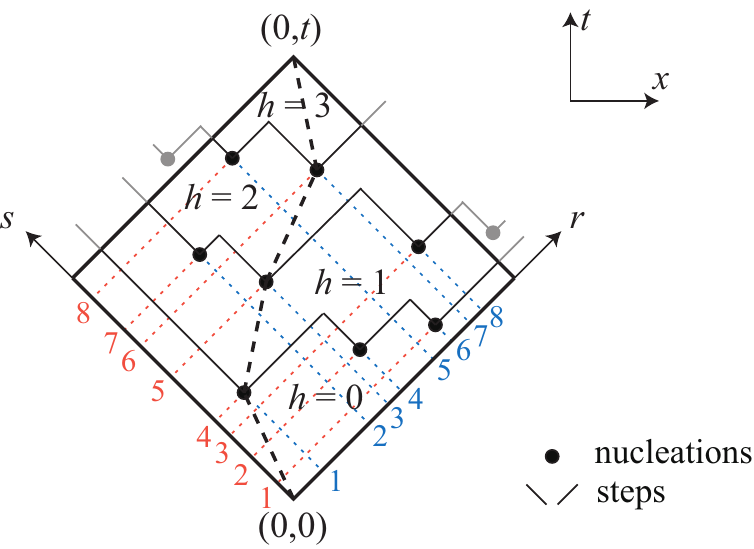}
 \caption{
An example of the space-time plot for the circular PNG problem.
This sketch is adapted from Fig.~1 in \cite{Prahofer.Spohn-PRL2000}.
}
 \label{fig5}
\end{figure}%

To describe the evolution of such a curved interface,
 it is useful to make a space-time plot
 showing where and when the nucleations occur,
 and along which paths their steps propagate.
Figure~\ref{fig5} shows an example of such a space-time plot.
In this plot, nucleations (black and gray dots) are generated randomly
 and independently at density $\rho = 1$ per unit area (Poisson point process),
 above the diagonal base lines that are depicted as the axes $r$ and $s$.
Each nucleation then produces steps (black and gray solid thin lines)
 that propagate diagonally.
When two diagonal lines meet, they terminate.

Now our quantity of interest is the height $h(0,t)$
 at position $x=0$ and time $t$.
First of all, because the ``speed of light'' of this PNG model is $\varv=1$,
 it is sufficient to consider only the square region
 enclosed by bold black lines in \figref{fig5}.
Next, we notice that the paths of the plateau steps (black thin lines)
 are the boundaries of the space-time regions with different heights.
In other words, the height $h(x,t)$ increases by one
 each time we cross a black thin line.
Therefore, 
\begin{equation}
 h(0,t) = \text{\# of black thin lines to cross when moving from point $(0,0)$ to point $(0,t)$,}  \label{eq:PNG1}
\end{equation}
 under the constraint that the trajectory is always oriented upward,
 or more precisely, within the angle $45^\circ$ to $135^\circ$.
Within the same constraint, it is also
\begin{equation}
 \pref{eq:PNG1} = \text{maximal \# of black dots passed by a string
 (=directed polymer) with end points fixed at $(0,0)$ and $(0,t)$,}
  \label{eq:PNG2}
\end{equation}
 as shown by the bold dashed line in \figref{fig5}.
In other words, it is the (sign-flipped) ground-state energy
 of this directed polymer under a random potential
 $U(x,t) = -\sum_n \delta(x-x_n)\delta(t-t_n)$.
Now we write down coordinates of the black dots, i.e., nucleation points.
Here the coordinates are simply defined as indices
 along the two diagonal axes $r$ and $s$, as illustrated in \figref{fig5}.
If we arrange the sets of the coordinates in ascending order
 along one of the two axes, say $r$, we obtain 
 $\begin{matrix} r: \\ s: \end{matrix} \begin{pmatrix} 1~2~3~4~5~6~7~8 \\ 4~7~5~2~8~1~3~6 \end{pmatrix}$ for the example of \figref{fig5}.
This is precisely a random permutation,
 whose length is generated randomly according to the Poisson distribution
 of mean $\rho S = t^2/2$, where $S$ is the area of interest
 (the square region).
Then, selecting a set of black dots to pass amounts to extracting a subsequence
 of the random permutation, and the directedness of the polymer
 is equivalent to the monotonic increase of the subsequence.
To sum up, together with Eqs.~\pref{eq:PNG1} and \pref{eq:PNG2},
\begin{multline}
 h(0,t) = \text{length of longest increasing subsequence of a random permutation} \\ \text{whose length is generated according to the Poisson distribution with mean $t^2/2$.}  \label{eq:PNG3}
\end{multline}
This is called Ulam's problem.

Now the problem is purely combinatorial.
In fact, there had been important advances in combinatorics
 \cite{Baik.etal-JAMS1999}
 that allowed one to explicitly compute this problem,
 by employing mathematical tools called the Young tableaux and
 the Robinson-Schensted correspondence.
The final result is surprisingly simple: for large $t$,
\begin{equation}
 h(0,t)
 \simeq 2\sqrt{S} + S^{1/6}\chi_2
 = \sqrt{2}t + (t/\sqrt{2})^{1/3} \chi_2.  \label{eq:PNGcirc}
\end{equation}
Here, the exponent $1/3$ confirms that this model is in the KPZ class,
 and $\chi_2$ is a random number generated according
 to a certain distribution known from random matrix theory
 \cite{Mehta-Book2004,Anderson.etal-Book2009},
 namely the GUE Tracy-Widom distribution \cite{Tracy.Widom-CMP1994}
 for the largest eigenvalue of GUE random matrices (see next subsection).
Moreover, it is straightforward to extend this result for $h(x,t)$,
 as it simply amounts to considering a rectangular region
 in the space-time plot, set by points $(0,0)$ and $(x,t)$.
We can use the same formula \pref{eq:PNGcirc} with $S = (t^2-x^2)/2$.
This also tells us that $\expct{h(x,t)} \simeq \sqrt{2(t^2-x^2)}$;
 in other words, our interface has a semicircle mean profile.

The bottom line so far is that the one-point distribution
 of the height fluctuations is explicitly obtained, is not Gaussian,
 and has a nontrivial connection to random matrix theory.
We have also seen that this model is related to a number of
 apparently different problems, such as directed polymer in random medium%
\footnote{
Note, however, that the connection between KPZ and the directed polymer
 has been known for a long time \cite{HalpinHealy.Zhang-PR1995},
 even from the beginning of the history \cite{Kardar.etal-PRL1986}.
}
 and Ulam's problem in combinatorics.

\subsection{Random matrix theory}  \label{sec:RMT}

Since a relation to random matrix theory has been encountered
 in the previous subsection,
 here we study some of its notions and results
 that are relevant for our purpose.

Random matrix theory \cite{Mehta-Book2004,Anderson.etal-Book2009}
 deals with fluctuation properties of eigenvalues of matrices
 made of random numbers.
Among them, matrices of Gaussian random numbers constitute
 the most fundamental ensembles of random matrices.
For example, an ensemble called the Gaussian unitary ensemble (GUE)
 considers a large complex Hermitian random matrix
\begin{equation}
 M = \begin{pmatrix}
 M_{11} & M_{12} & \cdots & M_{1N} \\
 M_{21} & M_{22} & \cdots & M_{2N} \\
 \vdots & \vdots & \ddots & \vdots \\
 M_{N1} & M_{N2} & \cdots & M_{NN}
 \end{pmatrix}  \label{eq:GUEmatrix}
\end{equation}
 with
\begin{equation}
 M_{ii} = a_{ii}, \qquad
 M_{ij} = \bar{M}_{ji} = a_{ij} + \ri b_{ij} \quad (i<j),
\end{equation}
 where $a_{ii} \in \mathbb{R}$
 is an independent and identically distributed (iid) Gaussian random number
 with mean $0$ and variance $1/2$,
 and $a_{ij}, b_{ij} \in \mathbb{R}$ with $i<j$ are iid Gaussian random numbers
 with mean $0$ and variance $1/4$.
This somewhat strange choice of the variances is made in order that
 the probability density of this matrix can be given
 by the following simple form%
\footnote{
It is an unfortunate conflict that the letter $\beta$ is used
 both for the parameter of the random matrix ensembles
 and for the growth exponent of kinetic roughening,
 predominantly in both contexts.
In these lecture notes, $\beta$ is used as the random matrix parameter
 only in \secref{sec:RMT};
 otherwise it is used as the interface growth exponent.
}
\begin{equation}
 P(M) = \frac{1}{Z} \e^{-\frac{\beta}{2}\Tr M^2},
 \label{eq:MatAPDF}
\end{equation}
 with $\beta \equiv 2$ and normalization constant $Z$.
This probability density is defined in such a way that $P(M) \rd M$ with
 $\rd M \equiv \prod_{i=1}^N \rd a_{ii} \prod_{i<j} \rd a_{ij} \rd b_{ij}$
 gives the probability to find the matrix elements $a_{ij}, b_{ij}$
 in the range $[a_{ij}, a_{ij}+\rd a_{ij})$ and $[b_{ij}, b_{ij}+\rd b_{ij})$.
Similarly, the Gaussian orthogonal ensemble (GOE) consists of
 real symmetric matrices, distributed according to \eqref{eq:MatAPDF}
 with $\beta = 1$,
 and the Gaussian symplectic ensemble (GSE) concerns
 quaternion self-dual Hermitian matrices with $\beta=4$.

Now we are interested in the distribution of the eigenvalues of the matrix $M$,
 denoted by $\lambda_1, \lambda_2, \cdots, \lambda_N$.
By diagonalizing, we have
\begin{equation}
 M = UXU^{-1}, \qquad
 X = \begin{pmatrix} \lambda_1 & & 0 \\ & \ddots & \\ 0 & & \lambda_N \end{pmatrix}, \qquad
 \rd M = J(U,X) \rd U \rd X,
\end{equation}
 where $U$ is an orthogonal (GOE), unitary (GUE), or symplectic (GSE) matrix
 and $J(U,X)$ is the Jacobian for this transformation.
By computing this Jacobian,
 one obtains the following probability density of the eigenvalues
 \cite{Mehta-Book2004,Anderson.etal-Book2009}
\begin{equation}
 P(\lambda_1, \cdots, \lambda_N) = \frac{1}{Z} \prod_{i<j} |\lambda_i - \lambda_j|^\beta \prod_i \e^{-\frac{\beta}{2}\lambda_i^2},  \label{eq:EigValPDF}
\end{equation}
 where we redefined the normalization $Z$.
This is equivalent to the canonical ensemble of particles $\lambda_i$
 interacting through the following Hamiltonian:
\begin{equation}
 H = \sum_i \frac{1}{2}\lambda_i^2 - \sum_{i<j} \log |\lambda_i - \lambda_j|,
\end{equation}
 i.e., with a harmonic trap and logarithmic repulsions,
 under the inverse temperature $\beta$.

We are almost there;
 what we encountered in the previous subsection was the distribution
 of the largest eigenvalue $\lambda_\mathrm{max}$ of GUE random matrices.
By using \eqref{eq:EigValPDF}, it is simply expressed by
\begin{equation}
 \Prob[ \lambda_\mathrm{max} \leq x ]
 = \Prob[ \lambda_1, \cdots, \lambda_N \leq x]
 = \frac{1}{Z} \int_{(-\infty,x]^N} \prod_i \rd \lambda_i \prod_{i<j} |\lambda_i - \lambda_j|^\beta \prod_i \e^{-\frac{\beta}{2}\lambda_i^2}.  \label{eq:LargestEigVal}
\end{equation}
For large $N$, it is known to scale as
\begin{equation}
 \lambda_\mathrm{max}
 \simeq \sqrt{2N} + 2^{-1/2}N^{-1/6} \chi_\mathrm{TW,\,\beta}.
 \label{eq:TWDef}
\end{equation}
The analytic expression for this $\chi_\mathrm{TW,\,\beta}$ was explicitly
 obtained by Tracy and Widom \cite{Tracy.Widom-CMP1994,Tracy.Widom-CMP1996},
 hence its distribution is called the Tracy-Widom distribution%
\footnote{
Now the Tracy-Widom distribution is popular enough,
 so that the function \texttt{TracyWidomDistribution}
 was implemented in Mathematica version 11.
Otherwise, a numerical table of the GUE and GOE Tracy-Widom distribution
 can be downloaded by courtesy of Pr\"ahofer and Spohn \cite{TracyWidomTable}.
}.
For reference, in the case of GUE, it is
\begin{equation}
 \Prob[\chi_\mathrm{TW,\,\beta=2} \leq x] \equiv F_2(x) = \det(1-P_x K P_x),
\end{equation}
 where $P_x$ is the projection onto $[x,\infty)$,
 $K(x,y)$ is the Airy kernel
\begin{equation}
 K(x,y) \equiv \frac{\Ai(x)\Ai'(y) - \Ai(y)\Ai'(x)}{x-y}
\end{equation}
 with the Airy function $\Ai(x)$,
 and $\det$ is the Fredholm determinant
 \cite{Mehta-Book2004,Anderson.etal-Book2009,Bornemann-MC2010},
 an extension of determinant for function space, defined by
\begin{equation}
 \det(1+zK) \equiv \sum_{n=0}^\infty \frac{z^n}{n!}\int_{(-\infty,\infty)^n} \det[K(x_i,x_j)_{i,j=1}^n] \rd x_1 \cdots \rd x_n.
\end{equation}
This can be approximated in terms of the usual matrix determinant
 \cite{Bornemann-MC2010}.
The GUE Tracy-Widom distribution can also be expressed as follows.
Let $u(x)$ be the global positive solution of the Painlev\'e II equation
 \cite{Conte.Musette-Book2008}
\begin{equation}
 \diffs{u}{x}{2} = 2u(x)^3 + xu(x),  \label{eq:Painleve2}
\end{equation}
 and define $g(x)$ such that $g''(x) = u(x)^2$
 and $g(x) \to 0$ as $x \to \infty$.
Then one simply has $F_2(x) = \e^{-g(x)}$
 \cite{Tracy.Widom-CMP1994,Prahofer.Spohn-PRL2000}.
With the same $g(x)$, and $f(x)$ such that $f'(x) = -u(x)$
 and $f(x) \to 0$ as $x \to \infty$, the GOE Tracy-Widom distribution
 $F_1(x) \equiv \Prob[\chi_\mathrm{TW,\,\beta=1} \leq x]$
 is given by $F_1(x) = \e^{-[f(x)+g(x)]/2}$
 \cite{Tracy.Widom-CMP1996,Prahofer.Spohn-PRL2000}.

\begin{figure}[t]
 \centering
 \includegraphics[scale=0.8,clip]{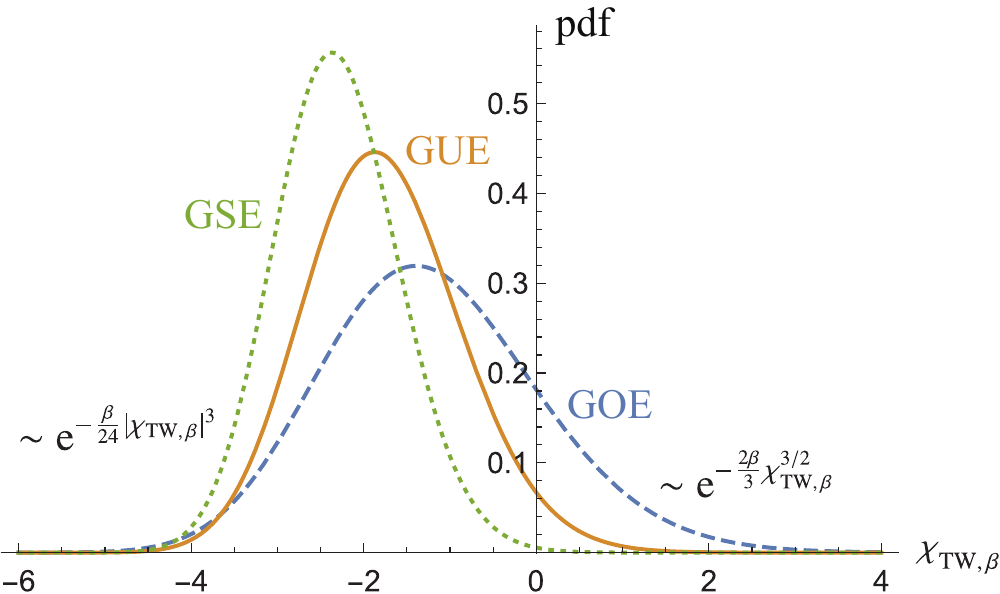}
 \caption{
Tracy-Widom distribution for GOE (blue dashed line), GUE (orange solid line),
 and GSE (green dotted line) random matrices. 
These were drawn by Mathematica \texttt{TracyWidomDistribution} function.
}
 \label{fig6}
\end{figure}%

Figure~\ref{fig6} shows the probability density
 of the Tracy-Widom distributions for GOE, GUE, and GSE random matrices.
As we can see, they are skewed and off-centered distributions.
Their left tail decays as $\e^{-\frac{\beta}{24}|\chi_\mathrm{TW,\,\beta}|^3}$
 and right tail as $\e^{-\frac{2\beta}{3}\chi_\mathrm{TW,\,\beta}^{3/2}}$
 \cite{Majumdar.Schehr-JSM2014}.
The left tail decays faster because,
 in order for the largest eigenvalue to be smaller,
 it needs to push all the other eigenvalues to its left.

\subsection{PNG flat interface}  \label{sec:PNGflat}

Now we come back to the exact results on the PNG model
 \cite{Prahofer.Spohn-PRL2000}.
In \secref{sec:PNGcirc}, we dealt with the PNG circular interfaces
 and the GUE Tracy-Widom distribution appeared
 as the one-point distribution of the height function,
 i.e., $\chi_2 = \chi_\mathrm{TW,2}$ in \eqref{eq:PNGcirc}.
In the present subsection, we consider flat PNG interfaces (\figref{fig4}(a)),
 obtained from the initial condition $h(x,0)=0$
 without any constraint on the nucleation region.

\begin{figure}[t]
 \centering
 \includegraphics[clip]{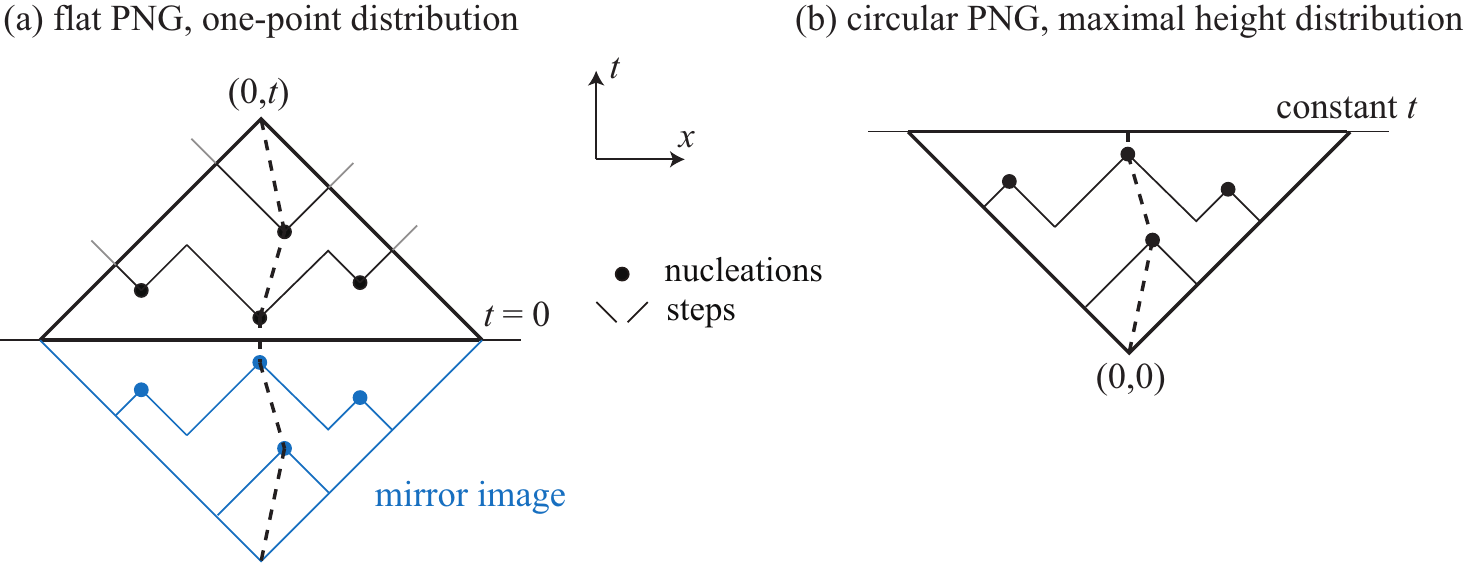}
 \caption{
An example of the space-time plot
 for the one-point distribution of the flat PNG problem (a)
 and for the maximal height distribution of the circular PNG problem (b).
}
 \label{fig7}
\end{figure}%

First we draw a space-time plot,
 similar to \figref{fig5} we did for the circular case.
In fact the only difference from the circular case
 is that now nucleations can occur anywhere on the substrate,
 so that the base line of the space-time plot is simply the horizontal line
 $t=0$ (\figref{fig7}(a))
 instead of the diagonal axes $r$ and $s$ for the circular case.
In other words, the space-time region to consider is a triangle
 with base on the line $t=0$ and top at the point $(0,t)$.
In terms of the directed polymer, it is now a ``point-to-line problem''
 (one fixed end and one free end), in contrast to the point-to-point problem
 (both ends fixed) for the circular case.
Then, by putting the mirror image beneath the line $t=0$,
 this triangular problem can be reduced to the squared problem
 we considered for the circular case,
 except that now we obviously have time-reversal symmetry
 in the nucleation points.
This symmetry is translated to
 a reflection symmetry of the random permutation%
\footnote{
Suppose an integer index $r$ (coordinate in \figref{fig5})
 is permuted to $s = p(r)$,
 this reflection symmetry reads $p(N+1-p(r)) = N+1-r$,
 where $N$ is the length of the permutation \cite{Prahofer.Spohn-PRL2000}.
},
 which is known to change the resulting distribution
 from the GUE Tracy-Widom to its GOE counterpart \cite{Baik.Rains-MSRIP2001}.
More specifically, since $h$ and $t$ in \figref{fig5} should be now replaced
 by $2h$ and $2t$, respectively, we obtain
\begin{align}
 &2h(0,t) \simeq \sqrt{2}(2t) + (2t/\sqrt{2})^{1/3}\chi_\mathrm{TW, 1},
 \notag \\
 &\therefore h(0,t) \simeq \sqrt{2}t + (t/\sqrt{2})^{1/3}\chi_1 \qquad
 \text{with $\chi_1 \equiv 2^{-2/3}\chi_\mathrm{TW, 1}$.}
\label{eq:PNGflat}
\end{align}
Therefore, apart from the uninteresting factor $2^{-2/3}$,
 the one-point distribution of the flat PNG interfaces
 is the GOE Tracy-Widom distribution, instead of the GUE Tracy-Widom
 that appeared for the circular case.
It is counterintuitive that fluctuation properties
 of the circular and flat interfaces are distinct
 even in the limit $t \to \infty$,
 because the mean profile of the circular interfaces
 becomes flatter and flatter as time goes on.
Nevertheless, it is the truth, at least for the PNG model,
 as it is a rigorous result.

For physicists, we may argue that
 the presence or absence of the time-reversal symmetry
 in the PNG space-time plot is essential for having the different asymptotic
 properties of the flat and circular interfaces.
This symmetry argument is indeed useful, because it helps to guess
 what we should obtain from the maximal height $\max_x h(x,t)$
 of the circular PNG interfaces.
Here, since the geometry is circular, we have the diagonal base lines
 in the space-time plot.
But now, at time $t$,
 we scan all end points $(x,t)$ with varying $x$
 to find the maximum of $h(x,t)$.
This corresponds to the triangular space-time geometry
 as drawn in \figref{fig7}(b), which has the exactly same symmetry
 as for the one-point distribution in the flat case.
We can therefore expect the GOE Tracy-Widom distribution
 for the maximal height of the circular interfaces,
 which was indeed shown to be true by both rigorous and non-rigorous approaches
 \cite{Johansson-CMP2003,Forrester.etal-NPB2011,Corwin.etal-CMP2013}.

\subsection{PNG stationary interface}  \label{sec:PNGstat}

\begin{figure}[t]
 \centering
 \includegraphics[clip]{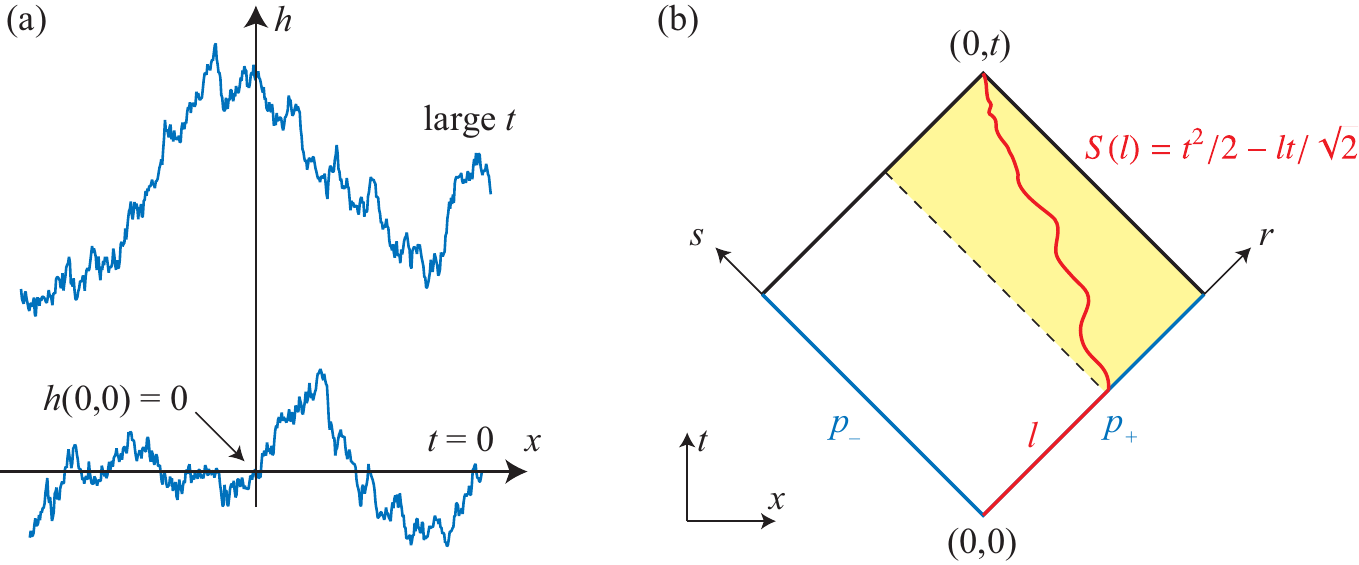}
 \caption{
Stationary PNG problem.
(a) Sketch of the stationary KPZ problem.
(b) Space-time plot to consider for the stationary PNG problem.
In addition to bulk nucleations at density $\rho$ per unit area,
 one also has boundary nucleations on the $r$- and $s$-axis,
 at density $p_+$ and $p_-$, respectively, per unit length.
As a result, the optimal path of the directed polymer (red bold line)
 can stay on the boundary for a finite length $l$
 and then wander in the bulk (yellow shaded area),
 whose remaining area is $S(l) = t^2/2 - lt/\sqrt{2}$.
}
 \label{fig8}
\end{figure}%

Finally we aim to investigate the (statistically) stationary state
 of the PNG model.
The question to ask is, if we take a stationary interface profile
 as the initial condition $h(x,0)$ and set the origin at $h(0,0)=0$,
 what the distribution of $h(0,t)$ is for large $t$
 (\figref{fig8}(a)).

To deal with this stationary PNG problem,
 we first consider the following extension of the circular problem:
 in addition to bulk nucleations that occur in $|x| \leq t$
 at rate $\rho=1$,
 we have nucleations at the boundaries $x = \pm t$,
 i.e., on the $r$- and $s$-axes of the space-time plot (\figref{fig8}(b)).
The boundary nucleations occur at rate $p_+$ and $p_-$
 on the $r$- and $s$-axes, respectively,
 per unit length of the space-time plot.
Depending on the balance between the bulk and boundary nucleations,
 the directed polymer can stay on either boundary over some length,
 denoted by $l$, before jumping to the bulk.
Once the polymer leaves the boundary, it cannot go back
 because of the directedness of the polymer.
Then the height $h(0,t)$ can be decomposed
 into the bulk and boundary contributions,
 $h_\mathrm{bulk}(l)$ and $h_\pm(l)$, respectively, as follows:
\begin{equation}
 h = \max_{l, \pm} (h_\mathrm{bulk}(l)+h_\pm(l))
\end{equation}
 with
\begin{equation}
 h_\mathrm{bulk}(l) \simeq 2\sqrt{S(l)} + S(l)^{1/6}\chi_2, \qquad
 h_\pm(l) \simeq (p_\pm l) + (p_\pm l)^{1/2}\chi_\mathrm{G},
\end{equation}
 where $S(l) = t^2/2 - lt/\sqrt{2}$ is the remaining area of the bulk
 (yellow shaded area of \figref{fig8}(b)),
 the double sign indicates the choice of the boundary,
 and $\chi_\mathrm{G}$ is a random variable drawn from the normal distribution
 (zero mean, unit variance).
To determine the maximizer $l$,
 for large $t$, it is sufficient to consider the leading terms only,
 $F(l) \equiv 2\sqrt{S(l)} + p_\pm l$.
From $F'(l_\mathrm{max})=0$, we obtain
\begin{equation}
 l_\mathrm{max} = \frac{t}{\sqrt{2}}\(1-\frac{1}{p_\pm^2}\).
\end{equation}
Therefore, if $p_\pm > 1$,
 the polymer stays on the boundary up to $l_\mathrm{max}$.
If $p_\pm < 1$, the polymer lives in the bulk only.
$p_\pm = 1$ is critical.
Assuming $p_+ \geq p_-$ without loss of generality,
 we can classify the situation as follows:
\begin{itemize}
\item
$p_- \leq p_+ < 1$:
 bulk is dominant, $h(0,t)$ shows the GUE Tracy-Widom distribution.
\item
$p_+ > 1$ and $p_+ \neq p_-$:
 boundary is dominant (for the fluctuations),
 $h(0,t)$ shows the Gaussian distribution.
\item
$p_+ = p_- > 1$:
 boundary is dominant, $h(0,t)$ shows the (Gaussian)$^2$ distribution,
 i.e., the distribution of the maximum of two iid Gaussian variables.
\item
$p_+=1$ and $p_- < 1$:
 critical, the distribution is known to be the (GOE Tracy-Widom)$^2$
 distribution \cite{Baik.Rains-JSP2000,Prahofer.Spohn-PRL2000}.
\item
$p_+ = p_- = 1$:
 critical, a new distribution was found for this case
  \cite{Baik.Rains-JSP2000,Prahofer.Spohn-PRL2000},
 called the Baik-Rains distribution.
\end{itemize}

Now we want to relate this external source problem
 to the stationary state of the PNG model.
Let $\rho_\stepup = \rho_\stepdown$ be the density
 of upward and downward steps, respectively, in the stationary state
 (steps of plateaux in \figref{fig4}).
In the stationary state, the step creation rate $\rho$ must be balanced
 with the step annihilation rate, so that
\begin{equation}
 \rho \rd x \rd t = (\rho_\stepup \rd x) (\rho_\stepdown 2v\rd t).
\end{equation}
Here the first factor of the right hand side indicates
 the probability to find an upward step in the range $\rd x$,
 and the second factor is the probability that it encounters
 a downward step within time interval $\rd t$.
Then we obtain $\rho_\stepup \rho_\stepdown = \rho/2v = 1/2$,
 hence $\rho_\stepup = \rho_\stepdown = 1/\sqrt{2}$,
 which is the density of steps measured along the $x$ axis.
If the density is measured along the $r$- and $s$-axes, it is one,
 which coincides exactly with $p_+ = p_- = 1$
 that gave the Baik-Rains distribution.
Therefore, for the PNG stationary state,
\begin{equation}
 h(0,t) \simeq \sqrt{2}t + (t/\sqrt{2})^{1/3} \chi_0,  \label{eq:PNGstat}
\end{equation}
 with $\chi_0$ being a random number drawn from the Baik-Rains distribution%
\footnote{\label{ft:StationaryHeight}
Note that $h(0,0) = 0$ by construction,
 while $h(x,0)$ is not necessarily zero for $x \neq 0$.
Therefore, for general $x$, one should measure the height difference in time,
 $h(x,t) - h(x,0)$, to find the Baik-Rains distribution.
}.

To the author's knowledge, no connection between the Baik-Rains distribution
 and random matrix theory is known.
However, using $f(x)$ and $g(x)$ associated with the Painlev\'e II equation
 \pref{eq:Painleve2}, the Baik-Rains distribution can be expressed as
 \cite{Baik.Rains-JSP2000,Prahofer.Spohn-PRL2000}
\begin{equation}
 \Prob[\chi_0 \leq x] \equiv F_0(x) = \[ 1-(x+2f''(x)+2g''(x))g'(x) \] \e^{-2f(x)-g(x)}.  \label{eq:BaikRains}
\end{equation}

\subsection{Universality subclass}

The results obtained
 in Secs.~\ref{sec:PNGcirc}, \ref{sec:PNGflat}, \ref{sec:PNGstat}
 prove that the one-point distribution of the PNG height fluctuations
 is different among the circular, flat, and stationary cases.
In other words, the distribution changes
 according to the global geometry of the interfaces,
 i.e., whether the interface is circular, flat, or stationary (Brownian),
 or equivalently, depending on the initial condition%
\footnote{
Instead of imposing the constraint on the nucleations
 in the circular PNG problem, one can equivalently use
 the following initial condition:
 $h(x,0) = 0$ at $x=0$ and $h(x,0)=-\infty$ otherwise.
}.
On the other hand, we may still expect universality%
\footnote{
Universality is expected from the viewpoint of physics of critical phenomena.
On the other hand, the exact derivation of the Tracy-Widom distributions
 relies on the integrability of the studied models,
 which is usually a fragile property.
Therefore, at least to the author, it is not trivial whether
 the Tracy-Widom distributions and related statistical properties arise
 universally in the KPZ class,
 including non-integrable models and experimental systems.
};
 the same distribution may arise for more general initial conditions,
 if they share the same symmetry (and other global properties),
 as well as for other systems in the KPZ class.
Indeed, there is now overwhelming evidence that
 this geometry dependence is not a special property of the PNG model;
 instead, it is known to occur in various models \cite{Kriecherbauer.Krug-JPA2010,Corwin-RMTA2012,HalpinHealy.Takeuchi-JSP2015,Quastel.Spohn-JSP2015,Sasamoto-PTEP2016} and in experiments \cite{Takeuchi.etal-SR2011,Takeuchi.Sano-JSP2012} as well.
Moreover, this is not restricted to the one-point distribution either;
 while the scaling exponents $\alpha, \beta, z$ are left unchanged,
 many other statistical properties, in particular correlation properties,
 are known to change accordingly
 (see \secref{sec:SpaceCorr} for the spatial correlation).
In view of this, it may be convenient to regard the three cases
 as different \textit{universality subclasses}, within the single KPZ class.
The different subclasses share the same scaling exponents,
 but are otherwise characterized by different statistical properties
 (\tblref{tbl:subclass}).
According to the usual belief, all subclasses are still controlled
 by the same KPZ fixed point of renormalization group theory.

\begin{table}[t]
 \begin{center}
  \caption{The three canonical subclasses of the 1D KPZ class.}
  \label{tbl:subclass}
  \catcode`?=\active \def?{\phantom{0}}
  {\small
  \begin{tabular}{l|ccc} \hline
  KPZ subclass & Circular & Flat & Stationary \\ \hline
  Standard initial condition & narrow-wedge (or point nucleus) & flat & Brownian \\
  & $h(x,0) = \begin{cases} 0 & (x=0) \\ -\infty & (x \neq 0) \end{cases}$ & $h(x,0)=0$ & $h(x,0) = \sqrt{A}B(x)$ \\
  Initial condition for ASEP & step (\figref{fig10}(a)) & alternating (\figref{fig11}(a)) & Bernoulli (\figref{fig11}(b)) \\ \hline
  Exponents & \multicolumn{3}{c}{$\alpha = 1/2, \beta = 1/3, z = 3/2$ for all subclasses} \\
  Distribution \cite{Prahofer.Spohn-PRL2000} & GUE Tracy-Widom ($\chi_2$) & GOE Tracy-Widom ($\chi_1$) & Baik-Rains ($\chi_0$) \\
 ~\footnotesize ($\expct{\chi_i}, \cum{\chi_i^2}, \Sk(\chi_i), \Ku(\chi_i)$)$^{\rm a}$ & \footnotesize (-1.77, 0.813, 0.224, 0.093) & \footnotesize (-0.760, 0.638, 0.293, 0.165) & \footnotesize (0, 1.15, 0.359, 0.289) \\
  Spatial process \cite{Quastel.Remenik-a2013} & Airy$_2$ process & Airy$_1$ process & Airy$_\mathrm{stat}$ process$^{\rm b}$ \\
  ~(its 2-pt correlation $g_i(\zeta)$) & (power-law decay) & (superexponential decay) & (superexponential decay) \\
  Two-time correlation & $F_\mathrm{t}(\tau) \sim \tau^{-1/3}$ & $F_\mathrm{t}(\tau) \sim \tau^{-1}$ & $F_\mathrm{t}(\tau) \sim \tau^{-1/3}$ \\
 ~\cite{Takeuchi.Sano-JSP2012,Ferrari.Spohn-SIG2016} & (ergodicity breaking) && (ergodicity breaking) \\
  Persistence exponents$^\mathrm{c}$ & $\theta_+ \approx 1.35, \theta_- \approx 1.85$ \cite{Takeuchi.Sano-JSP2012} & $\theta_\pm \approx 0.80$ \cite{Takeuchi.Sano-JSP2012} & $\theta_\pm = 1-\beta = 2/3$ \cite{Krug.etal-PRE1997,Kallabis.Krug-EL1999} \\ \hline
  \end{tabular}
  }
 \begin{spacing}{0.8}
 \raggedright\footnotesize $^{\rm a}$ $\cum{\chi_i^n}$ denotes the $n$th-order cumulant of $\chi_i$, $\Sk(\chi_i) \equiv \cum{\chi_i^3}/\cum{\chi_i^2}^{3/2}$ and $\Ku(\chi_i) \equiv \cum{\chi_i^4}/\cum{\chi_i^2}^{2}$. The values are cited from \cite{Prahofer.Spohn-PRL2000}, in which more precise values are given.\\
 \raggedright\footnotesize $^{\rm b}$ Note the different definition of the spatial process, \eqref{eq:StatLimitingProcess}.\\
 \raggedright\footnotesize $^{\rm c}$ The values of $\theta_+$ and $\theta_-$ are shown in the case of $\lambda>0$. They are exchanged otherwise \cite{Kallabis.Krug-EL1999}.
 \end{spacing}
 \end{center}
\end{table}

\subsection{Height rescaling}  \label{sec:Rescaling}

In order to generalize
 the results \pref{eq:PNGcirc} \pref{eq:PNGflat} \pref{eq:PNGstat}
 to other systems, we need to know
 how to rescale the height $h(x,t)$ and position $x$ as functions of $t$.
The procedure was established
 by Krug and coworkers \cite{Krug.Meakin-JPA1990,Krug.etal-PRA1992}.
First of all, it is known that all statistical quantities
 in the KPZ scaling limit can be rescaled by using two parameters,
 $A$ (as appeared in \eqref{eq:BrownianInitCond}) and $\lambda$.
$A$ can be estimated most straightforwardly from the height-difference
 correlation function
\begin{equation}
 C_\mathrm{h}(\ell,t) \equiv \expct{[h(x+\ell,t)-h(x,t)]^2} \simeq A\ell,
  \label{eq:ParamEstim1}
\end{equation}
 which amounts to measuring the mean squared displacement
 of the Brownian stationary interfaces, \eqref{eq:BrownianInitCond}.
In general, for large $t$, the height profile becomes locally Brownian,
 so that \eqref{eq:ParamEstim1} is valid up to the correlation length
 $\xi \sim t^{2/3}$.
For the other parameter $\lambda$, if one has control over the global tilt
 of the interface%
\footnote{
Numerically, one can use the shifted periodic boundary condition,
 $h(L,t) = h(0,t) + uL$.
} or knows the mean interface profile $\expct{h(x,t)}$,
 one can calculate the asymptotic growth speed
 $v_\infty(u) \equiv \lim_{t \to \infty}\expct{\prt{h}{t}}$
 as a function of the tilt $u \equiv \lim_{t \to \infty}\expct{\prt{h}{x}}$
 and use the following relation \cite{Krug.Meakin-JPA1990,Krug.etal-PRA1992}:
\begin{equation}
 \lambda = \left. \diffs{v_\infty(u)}{u}{2} \right|_{u=0}.
 \label{eq:ParamEstim2}
\end{equation}
In particular, if the interface grows isotropically in two-dimensional space
 at asymptotic speed $v_\infty$
 (as is the case in some experimental systems \cite{Takeuchi-JSM2014}),
 one simply has \cite{Takeuchi.Sano-PRL2010,Takeuchi.Sano-JSP2012}
\begin{equation}
 \lambda = v_\infty.  \label{eq:ParamEstim3}
\end{equation}
As an alternative method, if one can control the system size $L$
 and measure the asymptotic growth speed $v_\infty(L)$
 (in the direction normal to the interface) as a function of $L$,
 the following finite-size scaling can also be used
 \cite{Krug.Meakin-JPA1990,Krug.etal-PRA1992}:
\begin{equation}
 v_\infty(L) - \lim_{L\to\infty} v_\infty(L) = -\frac{\lambda A}{2L}.
  \label{eq:ParamEstim4}
\end{equation}

With the parameters $A$ and $\lambda$,
 the evolution of the height $h(x,t)$ can be expressed
 as follows for large $t$:
\begin{equation}
 h(x,t) \simeq v_\infty t + (\Gamma t)^{1/3}\chi(\zeta,t)  \label{eq:Height}
\end{equation}
 with
\begin{equation}
 \Gamma \equiv \frac{1}{2}A^2\lambda, \qquad
 \zeta \equiv \frac{x}{\xi(t)} \equiv \frac{x}{2(\Gamma t)^{2/3}/A}.
  \label{eq:Rescaling}
\end{equation}
The constant $v_\infty$ can be determined directly from \eqref{eq:Height},
 by using%
\footnote{
One can plot $\expct{\prt{h}{t}}$ against $t^{-2/3}$,
 linearly extrapolate to $t \to\infty$, and read the $y$-intercept.
}
\begin{equation}
 \Expct{\prt{h}{t}} \simeq v_\infty + (\const) \times t^{-2/3} \to v_\infty.
  \label{eq:VinfEstim}
\end{equation}
Then, Pr\"ahofer and Spohn \cite{Prahofer.Spohn-PRL2000} predicted that
 the statistical properties of the rescaled variable $\chi(\zeta,t)$
 are universal within the KPZ class,
 despite the different distribution functions
 obtained for the different subclasses: in the limit $t\to\infty$,
 $\chi(\zeta,t) \tod \chi_0, \chi_1, \chi_2$
 for the stationary, flat, and circular subclasses, respectively,
 where ``$\tod$'' denotes convergence in the distribution.

To test this prediction, one can define the rescaled height
\begin{equation}
 \hres(x,t) \equiv \frac{h(x,t) - v_\infty t}{(\Gamma t)^{1/3}}
 \simeq \chi(\zeta,t)  \label{eq:RescaledHeight}
\end{equation}
 and measure its cumulants: the mean $\expct{\hres}$,
 the variance $\cum{\hres^2} \equiv \expct{\delta \hres^2}$
 with $\delta \hres \equiv \hres-\expct{\hres}$,
 and the third- and fourth-order cumulants
 $\cum{\hres^3} \equiv \expct{\delta \hres^3}$
 and $\cum{\hres^4} \equiv \expct{\delta \hres^4} - 3\expct{\delta \hres^2}^2$,
 respectively.
In practice, the mean $\expct{\hres}$ often includes
 a strong finite-time correction in the order of $\mathcal{O}(t^{-1/3})$
 \cite{Takeuchi.Sano-JSP2012,Ferrari.Frings-JSP2011,Alves.etal-JSM2013},
 which stems from the next leading order to \eqref{eq:Height},
 $\mathcal{O}(t^0)$, and is often problematic.
To reduce this finite-time effect,
 one may try \cite{Fukai.Takeuchi-PRL2017}
\begin{equation}
 \expct{v_\mathrm{res}(x,t)} \equiv \frac{3t^{2/3}}{\Gamma^{1/3}}\(\Expct{\prt{h}{t}} - v_\infty\) \simeq \expct{\chi(\zeta,t)}.  \label{eq:RescaledHeight2}
\end{equation}
These rescaled cumulants can be directly compared
 with the known values of the cumulants
 \cite{Prahofer.Spohn-PRL2000}
 (\tblref{tbl:subclass})
 for the Baik-Rains ($\chi_0$, stationary), GOE Tracy-Widom
 ($\chi_1$, flat) and GUE Tracy-Widom distribution ($\chi_2$, circular)%
\footnote{
Here are some pitfalls:
(1) When comparing with the GOE Tracy-Widom distribution,
 be careful not to forget the factor $2^{-2/3}$ in \eqref{eq:PNGflat}.
(2) To study the stationary subclass, one should use $h(x,t) - h(x,0)$
 instead of $h(x,t)$ (see footnote~\ref{ft:StationaryHeight}).
(3) In this subsection we implicitly assume the translational symmetry in $x$.
For the isotropically growing circular interfaces, one may use
 the local radius as the definition of the height
 and formally retrieve the translational symmetry (see \secref{sec:Exp}).
} \footnote{\label{ft:ParamEstim}
In practice, however, the precision of the parameter estimation
 by Eqs.~\pref{eq:ParamEstim1}-\pref{eq:ParamEstim4} can be moderate,
 especially in experiments, due to finite-time effect, statistical errors,
 and/or limited control over the experimental conditions
 (such as the tilt $u$ and the system size $L$).
In such a case, one may use one of the cumulants
 to determine the parameter $\Gamma$,
 by setting that cumulant in the rescaled unit to be one,
 or alternatively the known value of the expected distribution. 
The price to pay is, of course, one then abandons
 testing the theoretical prediction for that particular cumulant,
 but still the prediction can be tested with the other cumulants
 and other statistical quantities (such as correlation functions).
It is not recommended to rescale the histogram in such a way that
 it has zero mean and unit variance, because it amounts to
 killing the two lowest-order cumulants; then
 the difference among $\chi_0, \chi_1, \chi_2$ are often too small
 to detect from noisy data \cite{Takeuchi-JSM2014}.
}.
Otherwise, amplitude ratios, such as skewness
 $\Sk(h) \equiv \cum{h^3}/\cum{h^2}^{3/2} \simeq \Sk(\chi)$
 and kurtosis $\Ku(h) \equiv \cum{h^4}/\cum{h^2}^2 \simeq \Ku(\chi)$,
 are free from parameter estimation error.
One may also try
\begin{equation}
 \frac{9t^2(\expct{\prt{h}{t}}-v_\infty)^2}{\cum{h^2}} = \frac{\expct{v_\mathrm{res}}^2}{\cum{\hres^2}} \simeq \frac{\expct{\chi}^2}{\cum{\chi^2}}. \label{eq:MeanVarianceRatio}
\end{equation}
The advantage of this quantity
 is that only the mean and the variance are involved
 (hence good statistical accuracy is expected),
 and only $v_\infty$ is needed, which may be estimated
 rather precisely by using \eqref{eq:VinfEstim}.

\section{Exact results on the 1D KPZ class 2: other important models and quantities}  \label{sec:Exact2}

\subsection{Asymmetric simple exclusion process}

\begin{figure}[t]
 \centering
 \includegraphics[clip]{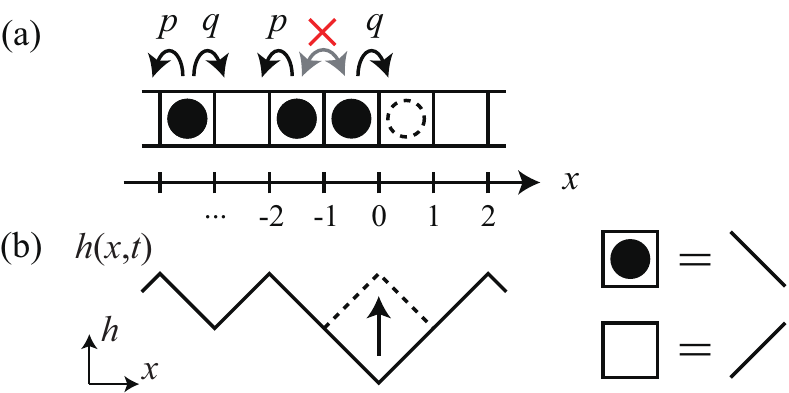}
 \caption{
Definition of the 1D ASEP (a) and mapping to a surface growth problem (b).
}
 \label{fig9}
\end{figure}%

The asymmetric simple exclusion process (ASEP),
 one of the paradigmatic models in non-equilibrium statistical physics
 \cite{Derrida-PR1998,Golinelli.Mallick-JPA2006},
 has also played a central role in the exact studies of the 1D KPZ class.
The model is defined on a lattice, here in 1D,
 with time $t \in \mathbb{R}$ and coordinate $x \in \mathbb{Z}$,
 which is assigned here to the midpoints of neighboring sites
 (\figref{fig9}(a)).
Each site is either occupied by a particle or empty.
Each particle can hop to a neighboring site stochastically and independently,
 on the right at rate $q$ and on the left at rate $p ~(<q)$,
 unless the new site is already occupied by another particle.
With $q>p$, the particles flow on average to the right.
The ASEP is usually regarded
 as a model of stochastic particle transport.
However, it can also be mapped to a model of interface growth,
 known by the name of single-step model in the literature,
 by replacing the occupied and empty sites with downward and upward slopes,
 respectively (\figref{fig9}(b)).
Then a forward jump of a particle corresponds to local increase
 of the interface at that point (compare two sketches of \figref{fig9}).
Therefore, the height increase, $h(x,t) - h(x,0)$,
 is given by the total (net) number of particles
 that have passed the position $x$ to the right.
In other words, $h(x,t)$ is the integrated current, which tends to grow
 because of $q>p$.
As expected from the evolution rule of the model,
 which consists of stochastic growth and local interactions
 (constraint that the interface slope is $\pm 1$),
 ASEP is indeed in the KPZ class.

\begin{figure}[t]
 \centering
 \includegraphics[clip]{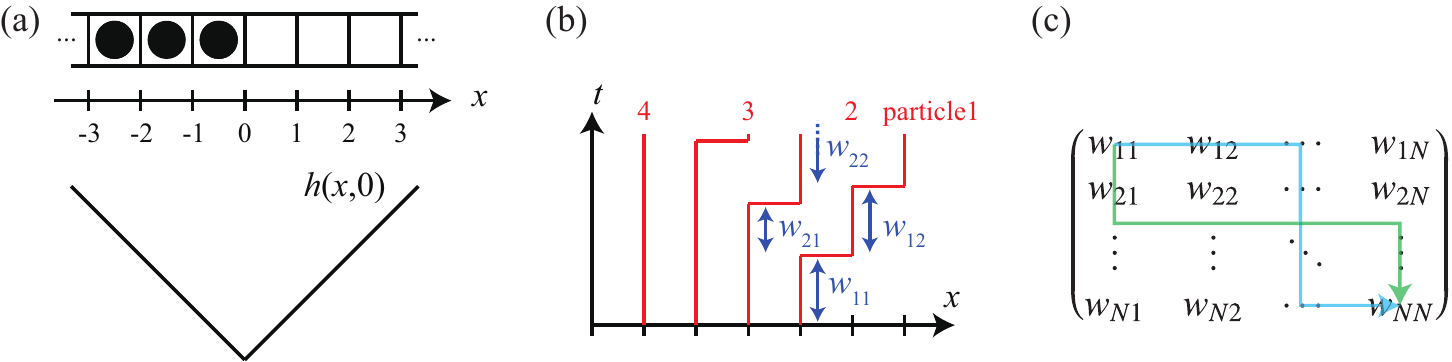}
 \caption{
Outline of the TASEP problem solved by Johansson \cite{Johansson-CMP2000}.
(a) Step initial condition and the corresponding initial interface profile.
(b) Trajectories of particles and waiting times $w_{ij}$.
(c) Matrix of $w_{ij}$ and last passage percolation.
}
 \label{fig10}
\end{figure}%

In fact, it is for the totally asymmetric version of ASEP (TASEP; $p=0$)
 that the height distribution was first solved exactly,
 by Johansson in 2000 \cite{Johansson-CMP2000}.
In the following we set $q=1$ without loss of generality%
\footnote{
Note that $q$ is the rate and not the probability;
 with $q=1$, each particle hops with probability $\rd t$ (if it is allowed)
 during an infinitesimal time step $\rd t$.
}.
We start from the step initial condition (\figref{fig10}(a)),
 where all sites with negative $x$ are occupied
 and all sites with positive $x$ are empty,
 and consider the number of particles that have passed $x=0$
 until time $t$, $N(t) ~(= h(0,t))$.
The quantity of interest is $\Prob[N(t) \geq N]$.
Then we notice that, in the case of TASEP,
 the proposition $N(t) \geq N$ is equivalent
 to the proposition $T_N \leq t$, where $T_N$ is the time at which
 the $N$th particle from the front makes the $N$th hopping.
This $T_N$ can be written in terms of waiting times $w_{ij}$
 for the $i$th particle to make the $j$th hopping.
If the waiting time $w_{ij}$ is counted
 only while the $i$th particle is able to hop,
 i.e., while the next site is empty
 (see \figref{fig10}(b)), $w_{ij}$ is simply an iid random number
 generated according to the exponential distribution, and $T_N$ is given by%
\footnote{
To understand \eqref{eq:TASEP1},
 the reader is recommended to see \figref{fig10}(b)
 and confirm (diagrammatically) that
 $T_2 = \max\{ w_{11} + w_{12} + w_{22}, w_{11} + w_{21} + w_{22}\}$.
}
\begin{equation}
 T_N = \max_{\pi \in \Pi} \sum_{(i,j) \in \pi} w_{ij}.  \label{eq:TASEP1}
\end{equation}
Here, $\Pi$ is a set of paths in the matrix of $w_{ij}$
 that start from the top left corner $w_{11}$, run only rightward or downward,
 and reach the bottom right corner $w_{NN}$ (\figref{fig10}(c)).
Equation~\pref{eq:TASEP1} is therefore a sort of optimization problem
 that maximizes the sum of $w_{ij}$ along such a directed path,
 called the last passage percolation problem.
We also notice that this is again a point-to-point problem of directed polymer
 (from the upper left to the lower right corner of the matrix $w_{ij}$)
 under random potential $w_{ij}$,
 akin to that considered for the circular PNG model
 (\figref{fig5} and \eqref{eq:PNG2}).

Johansson \cite{Johansson-CMP2000} solved this problem
 using combinatorial techniques,
 namely the Young tableaux and the Robinson-Schensted-Knuth correspondence,
 and reached the following equation:
\begin{equation}
 \Prob[N(t) \geq N] = \Prob[T_N \leq t] \propto \int_{[0,t]^N} \prod_{i=1}^N \rd x_i \prod_{i<j} (x_i-x_j)^2 \prod_i \e^{-x_i}.  \label{eq:TASEP2}
\end{equation}
Remarkably, this equation is very similar to that
 of the largest eigenvalue of GUE random matrices, 
 \eqref{eq:LargestEigVal} with $\beta=2$.
Indeed, the slight difference between the two equations
 turned out to be irrelevant; $N(t)$, or the height $h(0,t)$,
 was shown to be%
\footnote{
Johansson \cite{Johansson-CMP2000} obtained a formula for general $x$, 
 but for the sake of simplicity only the result for $x=0$ is shown here.
}
\begin{equation}
 h(0,t) = N(t) \simeq \frac{t}{4} - 2^{-4/3}t^{1/3}\chi_2,  \label{eq:TASEP3}
\end{equation}
 i.e., the height fluctuations show the GUE Tracy-Widom distribution.
This is the characteristic distribution of the circular KPZ subclass.
Why circular?
One may guess that this is because of the global curvature of the interface,
 associated with the step initial condition (see \figref{fig10}(a)).
But more appropriately, here the active region of the system,
 i.e., the region where particles exist and are not fully packed,
 expands in the course of time, just similarly to the nucleation region
 of the circular PNG problem (see \figref{fig4}(b)).
Interface fluctuations then do not propagate along the $h$-axis
 (line of constant $x$),
 but rather in the radial direction (radial when properly rescaled).
A related idea of direction of propagation was formulated mathematically, too,
 and called the characteristic line
 \cite{Ferrari-JSM2008,Corwin.etal-AIHPBPS2012}.

\begin{figure}[t]
 \centering
 \includegraphics[clip]{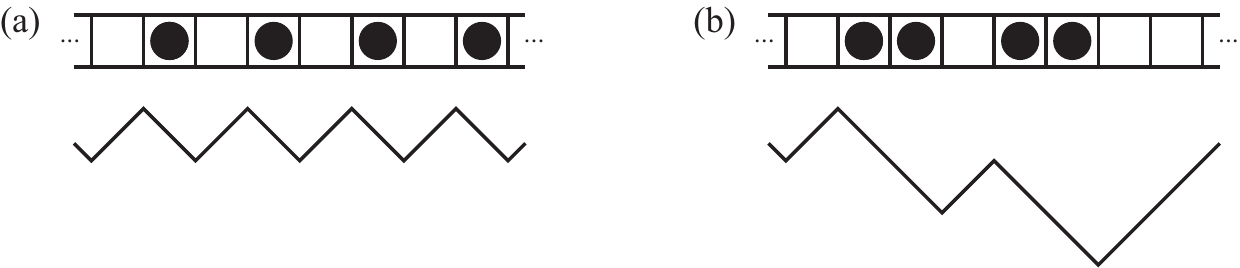}
 \caption{
Representative initial conditions considered in ASEP,
 other than the step initial condition for the circular subclass,
 already shown in \figref{fig10}(a).
(a) Alternating initial condition for the flat subclass.
(b) Bernoulli initial condition for the stationary subclass.
Here the particle configuration is generated randomly
 according to the Bernoulli measure; in other words,
 whether each site is occupied or empty is chosen randomly and independently
 with a fixed probability,
 $1/2$ in the case of the untilted stationary interface.
}
 \label{fig11}
\end{figure}%

With Johansson's result as the start, TASEP became
 one of the fundamental models to analyze
 rigorous properties of the 1D KPZ class.
To study the flat subclass, one can use the alternating initial condition
 where empty and occupied sites are alternating (\figref{fig11}(a)),
 and the GOE Tracy-Widom distribution was indeed derived
 \cite{Borodin.etal-JSP2007}.
The stationary subclass can also be studied
 with the Bernoulli initial condition, where the occupation of each site
 is determined randomly and independently with a fixed probability
 ($1/2$ to generate an untilted stationary interface; \figref{fig11}(b)),
 and the Baik-Rains distribution was identified \cite{Ferrari.Spohn-CMP2006}.
In contrast, solving the non-totally asymmetric case (i.e., $p \neq 0$)
 is much more difficult, because a property used in the mathematical results
 on TASEP, called free-fermionicity or complete determinantal structure,
 cannot be used.
However, it was finally solved for the step initial condition
 (circular subclass) by Tracy and Widom \cite{Tracy.Widom-CMP2009}
 via a Bethe ansatz approach,
 and for the Bernoulli initial condition (stationary subclass)
 as well \cite{Aggarwal-a2016,Aggarwal-MPAG2016}
 via relation to a variant of the six-vertex model.

From a different perspective, the type of optimization problems
 we considered in Eqs.~\pref{eq:PNG2} and \pref{eq:TASEP1}
 corresponds to the ground state of directed polymer,
 hence the zero-temperature problem.
Its finite-temperature analogue, i.e.,
\begin{equation}
 F = \frac{1}{\beta_T} \log Z, \qquad
 Z = \sum_{\pi \in \Pi} \exp\(-\beta_T \sum_{(i,j) \in \pi} E_{ij}\)
 = \sum_{\pi \in \Pi} \prod_{(i,j) \in \pi} \e^{-\beta_T E_{ij}}
 \label{eq:FiniteTempDP}
\end{equation}
 has also been considered and for some models
 the Tracy-Widom distribution was indeed shown \cite{Thiery.LeDoussal-JPA2015}.
Moreover, in some quantum interference problems,
 a summation similar to \eqref{eq:FiniteTempDP} can arise
 over forward scattering paths in the Green function
 \cite{Medina.Kardar-PRB1992},
 but then $\e^{-\beta_T E_{ij}}$
 is replaced by a not necessarily positive factor.
This arises for example in the context of conductance fluctuations
 of Anderson insulators, and interestingly the family of the Tracy-Widom
 distributions was still numerically observed
 \cite{Somoza.etal-PRL2007,Somoza.etal-PRB2015}.

\subsection{KPZ equation}

The series of rigorous approaches to the PNG model, ASEPs, and related models
 has finally led to exact solutions to the 1D KPZ equation
 (as defined in \eqref{eq:KPZeqHeightDef}
 with the 1D stochastic heat equation \pref{eq:SHE}), first in the year 2010%
\footnote{
Note that, being in the KPZ class does not necessarily indicate that
 the system is described by the KPZ equation,
 even on very long time scales and macroscopic length scales.
This can be seen by the fact that, for example,
 different models can have
 finite-time corrections (to the asymptotic distribution) of opposite signs,
 which cannot be changed by tuning the parameter values of the KPZ equation
 \cite{Ferrari.Frings-JSP2011}.
It was therefore important to prove whether the KPZ equation really shows
 the same asymptotic distribution as the solved discrete models.
}.
An important background was Tracy and Widom's solution
 of the (non-totally) ASEP \cite{Tracy.Widom-CMP2009},
 because the KPZ equation is actually known to be linked
 to certain weak asymmetric limit of ASEP
 \cite{Bertini.Giacomin-CMP1997,Sasamoto.Spohn-PRL2010,Sasamoto.Spohn-NPB2010,Amir.etal-CPAM2011}.
More specifically, set the hopping rates of ASEP to satisfy
 $q - p = a\sqrt{\ep}$ with asymmetry parameter
 $\ep \ll 1, a>0$, and $p + q = 1$, and denote the height profile of ASEP
 by $h_\mathrm{ASEP}^\ep(x,t)$ (as illustrated in \figref{fig9}).
Then, it was shown that
 $\sqrt{\ep}h_\mathrm{ASEP}^\ep(\floor{\ep^{-1}x},\ep^{-2}t)$
 in the limit $\ep\to 0$
 corresponds to the height of the KPZ equation
 (with $\nu=1/2$ and $\lambda = D = 1$) $h_\mathrm{KPZ}(x,t)$,
 up to some shifts.
The combination of this correspondence and
 the developments by Tracy and Widom on ASEP with the step initial condition,
 with other mathematical elaborations,
 led Sasamoto and Spohn
 \cite{Sasamoto.Spohn-PRL2010,Sasamoto.Spohn-NPB2010}
 and independently Amir \etal \cite{Amir.etal-CPAM2011}
 to establish the exact solution of the 1D KPZ equation
 for the narrow-wedge initial condition:
\begin{equation}
 h(x,0)
 = -\lim_{\delta\to 0^+} \frac{|x|}{\delta}
 = \begin{cases} 0 & (x=0), \\ -\infty & (x \neq 0), \end{cases} \qquad
 Z(x,0) \propto \delta(x).  \label{eq:KPZNarrowWedge}
\end{equation}
The exact solution gives an analytic expression
 for the one-point distribution of $h(x,t)$,
 which indeed converges to the GUE Tracy-Widom distribution
 of the circular KPZ subclass (in the rescaled unit).

Almost simultaneously, another approach based
 on the correspondence to the directed polymer problem
 and the use of the Bethe ansatz
 \cite{Franchini-Book2017}
 was established by Calabrese \etal \cite{Calabrese.etal-EL2010}
 and by Dotsenko \cite{Dotsenko-EL2010}.
To see this connection, let us first simplify the notation of the KPZ equation
 \pref{eq:KPZeq} by the following transformation
\begin{equation}
 x \mapsto \frac{(2\nu)^3}{D\lambda^2} x, \qquad
 t \mapsto \frac{(2\nu)^5}{D^2\lambda^4} t, \qquad
 h \mapsto \frac{2\nu}{\lambda}h.  \label{eq:Nondimensionalization}
\end{equation}
This amounts to setting $\nu=1/2$ and $\lambda = D = 1$ in the KPZ equation,
 without loss of generality,
 and the newly defined $x,t,h$ are dimensionless quantities.
Then we start from the stochastic heat equation \pref{eq:SHE}
 with smoothed noise $\eta_\kappa(x,t)$
 defined in Eqs.\pref{eq:SmoothNoise1} and \pref{eq:SmoothNoise2}
 (with $D=1$).
When converted to the Stratonovich form
 through Eqs.~\pref{eq:Ito2Stratonovich} and \pref{eq:StratonovichDivergence},
 it reads:
\begin{equation}
 \prt{}{t} Z(x,t) = \frac{1}{2} \vnabla^2 Z(x,t) + \eta_\kappa(x,t) \circ Z(x,t) - \frac{1}{2}\Delta_\kappa(0) Z(x,t).  \label{eq:SHEStratonivich}
\end{equation}
This has a form similar to the Schr\"odinger equation;
 then the solution $Z(x,t)$ of such an equation can be described
 by a path integral \`a la Feynman.
Indeed, according to the Feynman-Kac formula%
\footnote{
The Feynman-Kac formula tells us that the solution of
\begin{equation}
 \prt{Z}{t} = \nu\prts{Z}{x}{2} + U(x,t)Z(x,t)  \label{eq:FeynmanKac1}
\end{equation}
 with initial condition $Z(x,0) = \delta(x-x_0)$ is given by path integral
\begin{equation}
 Z(x,t) = \int_{B(0)=x_0}^{B(t)=x} \mathcal{D}B(\tau) \exp\left\{ -\int_0^t \rd \tau \[ \frac{1}{4\nu}\(\diff{B}{\tau}\)^2 - U(B(\tau),\tau) \]\right\},  \label{eq:FeynmanKac2}
\end{equation}
 where the two endpoints of the path $B(\tau)$ are fixed
 at $B(0)=x_0$ and $B(t)=x$.
If $U(x,t) = 0$,
 \eqref{eq:FeynmanKac1} is the diffusion equation,
 and the resulting path integral \pref{eq:FeynmanKac2} is simply
 the probability measure of Brownian motion;
 in other words, $B(\tau)$ can be regarded as Brownian motion,
 as naturally expected.
Now, with $U(x,t)$, the second term of \eqref{eq:FeynmanKac1} introduces
 exponential variation of the statistical weight $Z(x,t)$, which indeed appears
 in the action of \eqref{eq:FeynmanKac2}.
Of course, this is merely an interpretation;
 one should check the derivation of the formula to be convinced
 \cite{Majumdar-CS2005}.
} \cite{Majumdar-CS2005},
 the path integral expression of $Z(x,t)$ is given by
\begin{align}
 Z(x,t)
 &= \int \rd x_0 Z(x_0,0) \int_{B(0)=x_0}^{B(t)=x} \mathcal{D}B(\tau) \exp\left\{ -\int_0^t \rd \tau \[ \frac{1}{2}\(\diff{B}{\tau}\)^2 - \eta(B(\tau),\tau) + \frac{1}{2}\Delta_\kappa(0) \]\right\} \notag \\
 &= \int \rd x_0 Z(x_0,0) \int_{B(0)=x_0}^{B(t)=x} \mathcal{D}B(\tau) \e^{-E_\mathrm{DP}[B(\tau)]}  \label{eq:KPZDP}
\end{align}
 with
\begin{equation}
 E_\mathrm{DP}[B(\tau)] = \int_0^t \rd \tau \[ \frac{1}{2}\(\diff{B}{\tau}\)^2 - \eta(B(\tau),\tau) + \frac{1}{2}\Delta_\kappa(0) \].  \label{eq:KPZDPenergy}
\end{equation}
Here, the endpoints of the path $B(\tau)$ are fixed at $B(0)=x_0$ and $B(t)=x$
 in the path integral,
 but the first endpoint $x_0$ is distributed
 according to the probability density $\propto Z(x,0) = \e^{h(x,0)}$.
Importantly, \eqref{eq:KPZDP} can be interpreted as the partition function
 of a directed polymer $B(\tau)$ -- directed since time only increases --,
 with energy $E_\mathrm{DP}[B(\tau)]$ and temperature $k_\mathrm{B}T=1$.
According to \eqref{eq:KPZDPenergy}, its energy per unit length
 along the $\tau$-axis consists of
 elastic energy $\frac{1}{2}\(\diff{B}{\tau}\)^2$
 and random potential $-\eta(x,t)$
 (plus $\frac{1}{2}\Delta_\kappa(0)$).
The height $h(x,t) = \log Z(x,t)$ is then the (minus) free energy
 of the directed polymer.
Further, for the narrow-wedge initial condition \pref{eq:KPZNarrowWedge}
 (i.e., the circular subclass), the initial endpoint of the polymer is fixed
 at $(x,t)=(0,0)$; therefore, it is again the point-to-point problem.
For the flat case, the initial condition $Z(x,0) = \e^{h(x,0)}$ is constant, 
 so that the initial endpoint of the polymer is uniformly distributed
 along the line $t=0$.
This is the point-to-line problem, analogous to the flat case of the PNG model
 [\secref{sec:PNGflat} and \figref{fig7}(a)].

To investigate the fluctuations of $h(x,t)$, or equivalently $Z(x,t)$,
 one may examine the moments
 $\expct{Z(x_1,t)Z(x_2,t)\cdots Z(x_N,t)}$,
 which amounts to considering replicas of $Z(x,t)$.
Here the ensemble average is taken over noise realizations $\eta_\kappa(x,t)$.
Therefore, with $x_n(\tau)$ being the trajectory $B(\tau)$ of the $n$th replica
 and $\eta_n \equiv \eta_\kappa(x_n(\tau),\tau)$,
 the random potential contribution in \eqref{eq:KPZDP}
 during the time interval $[\tau, \tau+\rd\tau)$ 
 can be evaluated as follows:
\begin{equation}
 \Expct{\e^{\sum_n \eta_n \rd\tau}}
 = \int \(\prod_n \rd\eta_n \) \e^{\sum_n \eta_n \rd\tau} \exp\(\sum_{n,m} \frac{\eta_n\eta_m}{2\Delta_\kappa(x_n - x_m)/\rd\tau}\)
 \propto \exp\(\frac{1}{2}\sum_{n,m}\Delta_\kappa(x_n - x_m) \rd\tau\),
\end{equation}
 where the right hand side is obtained by the Gauss integral over $\eta_n$.
Then we have
\begin{align}
 &\expct{Z(x_1,t)Z(x_2,t)\cdots Z(x_N,t)} \notag \\
 &\qquad \propto \int \(\prod_n \rd x_{n0} Z(x_{n0},0)\) \int_{x_n(0)=x_{n0}}^{x_n(t)=x_n} \(\prod_n \mathcal{D}x_n(\tau)\) \exp\left\{ -\int_0^t \rd\tau \[\frac{1}{2}\sum_n \(\diff{x_n}{\tau}\)^2 - \frac{1}{2}\sum_{n \neq m}\Delta_\kappa(x_n - x_m) \]\right\},  \label{eq:KPZcumulant1}
\end{align}
 where trajectories $x_n(\tau)$
 are initially distributed according to $Z(x,0)$
 (expressed by the first integral)
 and arrive at $x_n(t) = x_n$.
Now we switch to the quantum mechanics representation.
By applying the Feynman-Kac formula to this many-body problem,
 now from the path-integral representation \pref{eq:KPZcumulant1}
 back to the partial differential equation,
 we obtain the following Schr\"odinger-like equation
\begin{equation}
 \prt{}{t}\Psi(\vx,t) = \[\frac{1}{2} \sum_n \prts{}{x_n}{2} + \frac{1}{2}\sum_{n \neq m}\Delta_\kappa(x_n - x_m)\] \Psi(\vx,t) \equiv -\hat{H}_\kappa \Psi(\vx,t)
\end{equation}
 with $\vx \equiv (x_1, \cdots, x_N)$ and Hamiltonian
\begin{equation}
 \hat{H}_\kappa \equiv -\frac{1}{2} \sum_n \prts{}{x_n}{2} - \frac{1}{2}\sum_{n \neq m}\Delta_\kappa(x_n - x_m).
\end{equation}
The KPZ equation is retrieved by $\kappa \to 0$.
Therefore, the corresponding Hamiltonian is
\begin{equation}
 \hat{H}_\mathrm{LL} = -\frac{1}{2} \sum_n \prts{}{x_n}{2} - \frac{1}{2}\sum_{n \neq m}\delta(x_n - x_m)  \label{eq:LL}
\end{equation}
 and the cumulants can be expressed as follows
 by using the quantum mechanics representation:
\begin{equation}
 \expct{Z(x_1,t)Z(x_2,t)\cdots Z(x_N,t)} = \braket{\vx}{\e^{-\hat{H}_\mathrm{LL}t}}{\vx_0}, \label{eq:KPZcumulantLL}
\end{equation}
 where $\ket{\vx_0}$ represents the initial condition
 ($\ket{\bm{0}}$ for the narrow-wedge case).
The Hamiltonian \pref{eq:LL} describes bosons with contact interactions,
 known as (the attractive version of)
 the Lieb-Liniger model \cite{Lieb.Liniger-PR1963}.
It is a well-known integrable system \cite{Franchini-Book2017},
 which has also an experimental relevance
 as recently demonstrated by experiments on the Bose-Einstein condensation
 of cold atom gas for the repulsive case
 \cite{Fabbri.etal-PRA2015,Meinert.etal-PRL2015}.
It is interesting that the KPZ problem is linked
 to such an apparently distinct problem.

From the theoretical viewpoint, this connection to the Lieb-Liniger model is
 important, because one can employ the Bethe ansatz \cite{Franchini-Book2017}
 to carry out the computation of the cumulants $\cum{Z(0,t)^n}$
 from \eqref{eq:KPZcumulantLL}
 \cite{Calabrese.etal-EL2010,Dotsenko-EL2010,Calabrese.LeDoussal-PRL2011,Imamura.Sasamoto-PRL2012}.
Although in general the conversion from cumulants to distribution
 is not unique, and even worse in the present problem
 a rapidly diverging term appears, which cannot be treated
 with mathematical rigor, this approach successfully gave
 the GUE Tracy-Widom distribution for the narrow-wedge initial condition
 \cite{Calabrese.etal-EL2010,Dotsenko-EL2010}.
This method also seems to be versatile;
 it was adapted to other initial conditions as well, and indeed,
 the GOE Tracy-Widom distribution was found for the flat case
 $h(x,0) = 0$ \cite{Calabrese.LeDoussal-PRL2011}
 and the Baik-Rains distribution for the stationary initial condition
 $h(x,0) = \sqrt{A}B(x)$ (\eqref{eq:BrownianInitCond}) with $h(0,0)=0$
 \cite{Imamura.Sasamoto-PRL2012}.
The stationary result was also proved mathematically later
 \cite{Borodin.etal-MPAG2015}.

Now, there are a number of mathematical approaches to study the KPZ equation
 (a nice summary can be found in introduction of \cite{Borodin.etal-MPAG2015}),
 which have deepen the mathematical understanding of this equation even more.
In the meantime, the replica Bethe ansatz approach continues being
 a powerful method of theoretical physics, which allows us to compute
 a variety of statistical quantities.

\subsection{Spatial correlation}  \label{sec:SpaceCorr}

So far we have focused on the one-point distribution of $h(x,t)$,
 but it is by no means the only quantity
 studied in recent developments on exact solutions.
Among others, one of the most important statistical properties
 that have been deeply understood is the spatial correlation,
 e.g., joint distribution of $h(x,t)$ at multiple points
 $x_1, x_2, \cdots$ at a single time $t$.
It is usually described by the rescaled coordinates,
 in other words, in terms of the rescaled height profile $\chi(\zeta,t)$
 defined by \eqref{eq:Height}.
In the limit $t\to\infty$, the joint distribution of $\chi(\zeta,t)$
 becomes independent of $t$. 
Consequently, $\chi(\zeta,\infty)$ can be regarded
 as a ``stochastic process'', with $\zeta$ playing the role of time
 for this process.
It is analogous to the way
 the stationary interface profile of the 1D KPZ equation was described
 by a Brownian motion in \eqref{eq:BrownianInitCond}.
Then the spatial correlation properties of KPZ-class interfaces
 can be expressed as the time correlation properties
 of such stochastic processes.

Analytic expressions for such limiting processes have been obtained
 for the the circular and flat subclasses \cite{Quastel.Remenik-a2013},
 and named the Airy$_2$ process $\mathcal{A}_2(\zeta)$ for the circular case
 \cite{Prahofer.Spohn-JSP2002,Johansson-CMP2003} and 
 the Airy$_1$ process $\mathcal{A}_1(\zeta)$
 for the flat case \cite{Sasamoto-JPA2005,Borodin.etal-JSP2007}.
More specifically, $\chi(\zeta,t)$ is predicted to converge as follows:
\begin{equation}
 \chi(\zeta,t) \tod \begin{cases} \mathcal{A}_2(\zeta) - \zeta^2 & \text{(circular subclass)} \\ \mathcal{A}_1(\zeta) & \text{(flat subclass)}
 \end{cases} \qquad (t \to \infty)  \label{eq:AiryProcessConv}
\end{equation}
 where ``$\tod$'' now denotes convergence of multi-point distributions.
The term $-\zeta^2$ in the circular case comes from the curvature
 of the circular interfaces%
\footnote{\label{ft:Airy1Def}
In relation to the factor $2^{-2/3}$
 between $\chi_1$ and $\chi_{\mathrm{TW}, 1}$ in \eqref{eq:PNGflat},
 the Airy$_1$ process $\mathcal{A}_1(\zeta)$ can be normalized differently,
 in such a way that $\chi(\zeta,t)$ converges
 to $2^{1/3}\mathcal{A}_1(2^{-2/3}\zeta)$
 \cite{Quastel.Remenik-a2013,Corwin.etal-CMP2013}.
In these lecture notes, however, we adopt the normalization that satisfies
 \eqref{eq:AiryProcessConv}, which is simpler in the KPZ context.
}.
As such, the one-point distribution of $\mathcal{A}_i(\zeta)$
 is simply the distribution of $\chi_i$, namely the GUE and GOE Tracy-Widom
 distributions ($i=2$ and $1$, respectively).
For the stationary subclass, a similarly defined limiting process would be just
 the Brownian motion (\eqref{eq:BrownianInitCond});
 therefore, one can consider instead \cite{Prahofer.Spohn-JSP2004}
\begin{equation}
 \chi_\mathrm{stat}(\zeta,t) \equiv (\Gamma t)^{-2/3}\expct{(h(x,t)-h(0,0)-v_\infty t)^2}.  \label{eq:StatLimitingProcess}
\end{equation}
For this object, the limiting process $\mathcal{A}_\mathrm{stat}(\zeta)$,
 named the Airy$_\mathrm{stat}$ process,
 was indeed formulated \cite{Baik.etal-CPAM2010}.

Interestingly, a direct connection between the Airy$_2$ process 
 and GUE random matrices is also known: the former is actually equivalent
 to a time-dependent process of a GUE random matrix
 \cite{Johansson-CMP2003,Quastel.Remenik-a2013},
 called Dyson's Brownian motion.
In this model, each element of an $N \times N$
 GUE matrix $M(t)$ (\eqref{eq:GUEmatrix})
 is assumed to perform an independent Ornstein-Uhlenbeck process
 \cite{Mehta-Book2004,Anderson.etal-Book2009}, that is,
\begin{equation}
 \diff{M(t)}{t} = -M(t) + \Xi(t),  \label{eq:DysonBM}
\end{equation}
 where $\Xi(t)$ is a Hermitian matrix
 with white Gaussian noise elements $\Xi_{ij}(t)$,
 whose mean is $\expct{\Xi_{ij}(t)}=0$ and covariance is
 $\expct{\Xi_{ii}(t)\Xi_{i'i'}(t)} = \delta_{ii'}\delta(t-t')$
 for the diagonal elements and
 $\expct{[\re\Xi_{ij}(t)][\re\Xi_{i'j'}(t)]} = \expct{[\im\Xi_{ij}(t)][\im\Xi_{i'j'}(t)]} = (1/2)\delta_{ii'}\delta_{jj'} \delta(t-t')$
 for the non-diagonal elements ($i<j, i'<j'$).
Then we focus on its largest eigenvalue $\lambda_\mathrm{max}(t)$.
Analogously to the one-point rescaling given in \eqref{eq:TWDef},
 $\lambda_\mathrm{max}(t)$ was shown to converge to the Airy$_2$ process
 \cite{Johansson-CMP2003,Quastel.Remenik-a2013} as
\begin{equation}
 2^{1/2}N^{1/6} \[\lambda_\mathrm{max}(\zeta/N^{1/3}) - \sqrt{2N}\]
 \tod \mathcal{A}_2(\zeta).  \label{eq:Airy2DysonBM}
\end{equation}
One might also expect a similar relationship between
 the Airy$_1$ process and Dyson's Brownian motion for GOE random matrices,
 but curiously the two processes are not the same
 \cite{Bornemann.etal-JSP2008}.
In other words, the mathematical structure behind the 1D KPZ class
 does \textit{not} overlap entirely with random matrix theory.

From a practical viewpoint, a useful quantity to characterize
 the spatial correlation is the two-point correlation function,
 defined by
\begin{equation}
C_\mathrm{s}(\ell,t) \equiv \cum{h(x+\ell,t)h(x,t)} \equiv \expct{\delta h(x+\ell,t) \delta h(x,t)}  \label{eq:SpaceCorr}
\end{equation}
 with $\delta h(x,t) \equiv h(x,t) - \expct{h(x,t)}$.
According to Eqs.~\pref{eq:Height} and \pref{eq:Rescaling},
 the correlation function may be rescaled as follows
 and is expected to converge
 to the two-point correlation of the corresponding Airy process%
\footnote{
According to \cite{Bornemann.etal-JSP2008},
 a numerical table of $g_1(\zeta)$ and $g_2(\zeta)$
 can be obtained from Bornemann by his courtesy.
Be aware of the different definition of $\mathcal{A}_1(\zeta)$
 as stated in footnote~\ref{ft:Airy1Def}.
}:
\begin{equation}
 C_\mathrm{s,res}(\zeta,t)
 \equiv (\Gamma t)^{-2/3} C_\mathrm{s}(\zeta\cdot\xi(t),t)
 \to g_i(\zeta) \equiv \cum{A_i(\zeta)A_i(0)}.
  \label{eq:SpaceCorrRescaling}
\end{equation}
Strikingly, the Airy correlation functions $g_i(\zeta)$ are known to decay
 in \textit{qualitatively} different manners:
 while the spatial correlation of the circular subclass decays
 by a power law, $g_2(\zeta) \sim \zeta^{-2}$
 \cite{Adler.Moerbeke-AP2005,Bornemann.etal-JSP2008},
 that of the flat subclass decays superexponentially,
 specifically, $g_1(\zeta) \sim \e^{-c_1|\zeta|^3}$ with constant $c_1$
 \cite{Ferrari-PC2017,Bornemann.etal-JSP2008}.
This is in contrast to the fact that
 the GUE and GOE Tracy-Widom distributions are only quantitatively different
 (see \figref{fig6}).
For the stationary subclass, too, the two-point function defined analogously
 decays superexponentially,
 $g_0(\zeta) \sim \e^{-c_0|\zeta|^3}$
 with a coefficient $c_0$ different from $c_1$
  \cite{Prahofer.Spohn-JSP2004,Ferrari-PC2017}.

In contrast to spatial correlations, correlation properties in time
 have remained an analytically challenging problem,
 but a few results started to emerge very recently
 \cite{Dotsenko-JSM2013,Ferrari.Spohn-SIG2016,Johansson-CMP2017,DeNardis.etal-PRL2017,Nardis.Doussal-JSM2017}.
We shall come back to this point in \secref{sec:ExpTimeCorr}.

\section{Liquid crystal experiment}  \label{sec:Exp}

While the recent developments on the 1D KPZ problem have been mostly
 driven by mathematical approaches, the obtained results have
 striking physical implications and the quantities of interest 
 are, at least in principle, observable in experiments and simulations.
Testing universality of the wealth of mathematical results is crucial,
 because essentially all those developments rely on the integrability
 of the studied models, which constitute only a special group
 of systems in the large KPZ universality class.
Experimentally, as already noted in footnote~\ref{ft:Experiments},
 even the scaling exponents of the KPZ class
 have been rarely found \cite{Takeuchi-JSM2014},
 the first examples being growing interfaces of mutant bacteria colony
 \cite{Wakita.etal-JPSJ1997} and slow combustion of paper
 \cite{Maunuksela.etal-PRL1997,Myllys.etal-PRE2001}.
Recently, however, a growing number of experiments have been reported,
 in which the two independent exponents of the 1D KPZ class
 were clearly observed \cite{Takeuchi-JSM2014}.
Here we focus on one such experiment, namely that on growing interfaces
 in liquid crystal turbulence conducted by Takeuchi and Sano
 \cite{Takeuchi.Sano-PRL2010,Takeuchi.etal-SR2011,Takeuchi.Sano-JSP2012}.
This experiment was able to provide sufficient accuracy
 to study fine statistical quantities,
 such as the distribution and correlation functions.
For the description of other experimental systems, the readers
 are kindly referred to the author's recent review \cite{Takeuchi-JSM2014}
 and references therein.

\subsection{Experimental system}

\begin{figure}[t]
 \centering
 \includegraphics[clip]{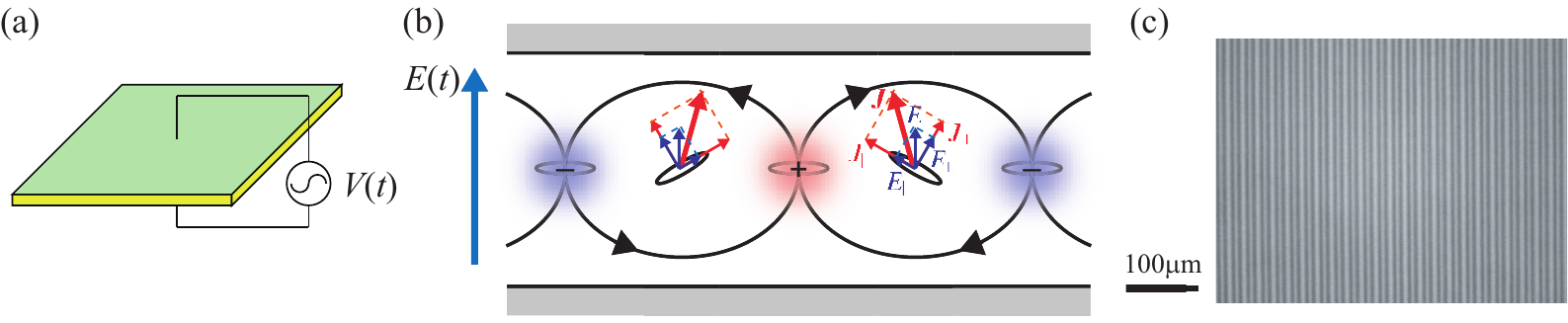}
 \caption{
Sketch of the experimental system of liquid crystal electroconvection (a)
 and the Carr-Helfrich mechanism for roll convection (b),
 as well as an actual microscope image of the roll convection (c).
(a) The system consists of nematic liquid crystal
 between two parallel glass plates, coated with transparent electrodes.
The gap is typically 10 to 50~$\mu$m and the span-wise size is about 1~cm.
For the experiment described in these notes
 \cite{Takeuchi.etal-SR2011,Takeuchi.Sano-JSP2012},
 the gap is roughly 12~$\mu$m and the span-wise size is 16~mm.
(b) This sketch shows a side view of the system.
In the planar alignment condition and in the absence of field,
 liquid crystal molecules in bulk are also aligned
 in a direction parallel to the surfaces.
Now, under a weak ac field $E(t)$, we consider a small perturbation
 of the director field as sketched (by the tilted ellipsoids),
 which is taken to be a single Fourier component.
Since the director field is now not perpendicular to the electric field, 
 to use Ohm's law, we need to decompose the electric field into
 the components parallel and normal to the director field,
 $E_\parallel$ and $E_\perp$,
 respectively, and apply Ohm's law to each component.
Because of $\sigma_\parallel > \sigma_\perp$,
 the induced current $\bm{J}$ has a lateral component,
 which leads to spatially periodic concentration
 of positive and negative charges.
Those charges are also subjected to the field,
 positive changes forced upward and negative ones downward,
 hence generating circular flow as shown in the panel.
This flow exerts a torque to the molecules,
 in the direction that amplifies the considered perturbation.
There is also an effect from the anisotropic dielectric constants
 $\ep_\parallel < \ep_\perp$, which also favors the deformed director field.
The amplitude of the deformation is determined
 by the balance of those effects and the nematic elasticity.
Since the electric field is ac, its direction oscillates periodically.
At low frequencies considered here, the currents and charges are oscillating
 with the external field, while the flow and the director field remain static.
See \cite{deGennes.Prost-Book1995} for more detailed descriptions
 of the mechanism described here.
(c) A microscope image of the roll convection.
The anisotropic refractive index of liquid crystal allows us to observe
 the convection pattern by the simple bright-field microscopy
 (i.e., by the transmitted light) \cite{deGennes.Prost-Book1995},
 without the need for tracer particles.
}
 \label{fig12}
\end{figure}%

\begin{figure}[t]
 \centering
 \includegraphics[scale=0.8,clip]{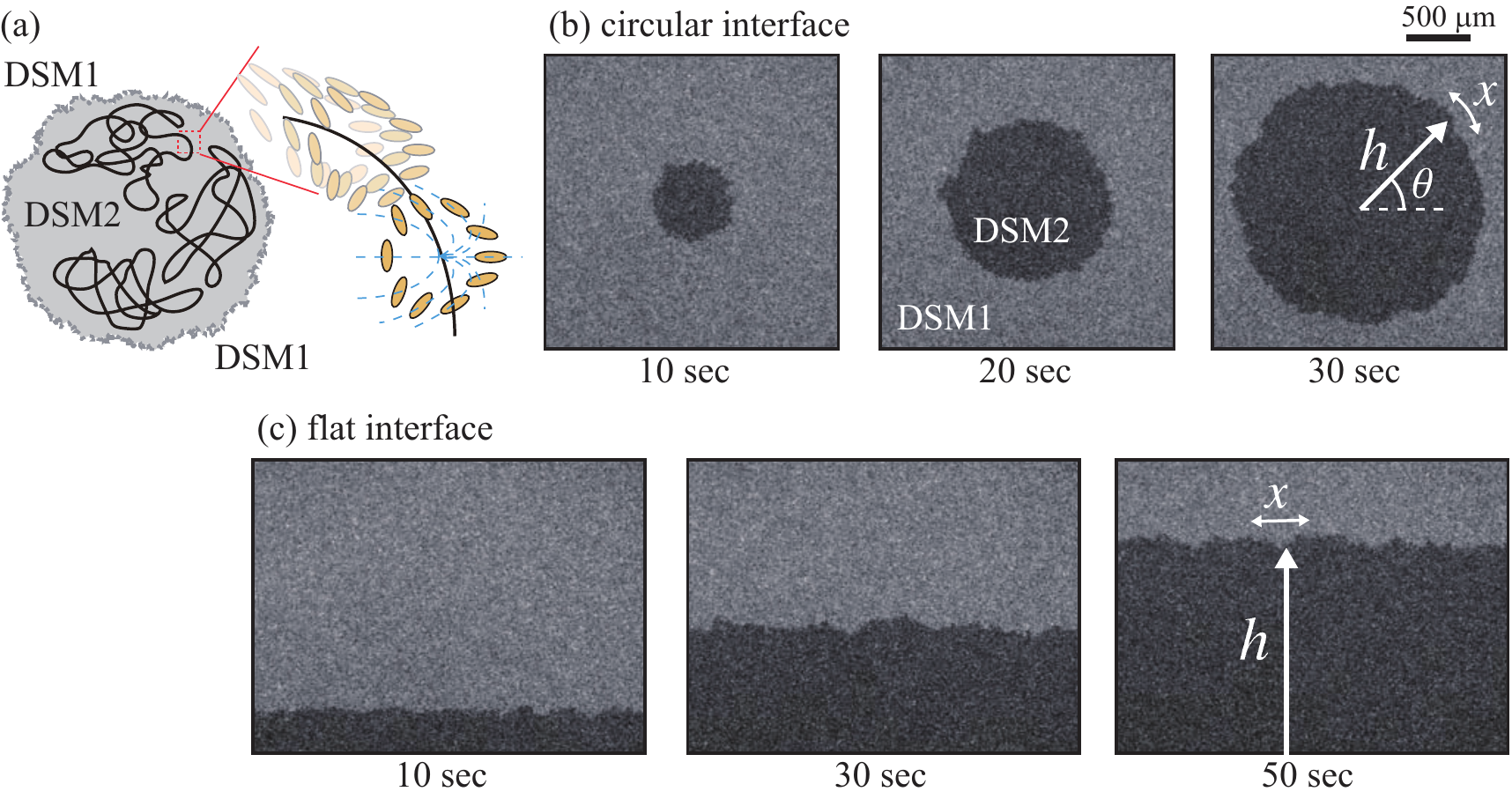}
 \caption{
(a) Sketch of the DSM2 state which consists of topological defects,
 specifically disclination lines, in the nematic director field.
Such defects are absent in the DSM1 state.
(b,c) Snapshots of growing DSM2 clusters under homeotropic alignment.
The growth was triggered by shooting laser pulses,
 either to a point or to a line, hence generating
 circular (b) and flat (c) interfaces, respectively
 \cite{Takeuchi.Sano-JSP2012}.
Time is measured from the emission of laser pulses.
Reprints with permission from \cite{Takeuchi.Sano-JSP2012}
 with some adaptations.
}
 \label{fig13}
\end{figure}

The system to study is electroconvection of nematic liquid crystal
 \cite{deGennes.Prost-Book1995}, i.e., convection driven by an electric field
 applied between two parallel glass plates (\figref{fig12}(a)).
In the standard setup, the surface alignment of liquid crystal
 is set to be parallel to the glass plates (planar alignment).
Electroconvection is observed if the liquid crystal
 has the following type of anisotropy in material properties:
 $\ep_\parallel < \ep_\perp$ in the dielectric constants
 and $\sigma_\parallel > \sigma_\perp$ in the conductivity,
 where $\parallel$ and $\perp$ indicate
 the directions with respect to the alignment of liquid crystal,
 or the director field.
The mechanism is described by the Carr-Helfrich theory
 \cite{deGennes.Prost-Book1995} in the case of the weak ac field
 at relatively low frequencies,
 as illustrated in \figref{fig12}(b) with explanations in its caption.
Increasing the applied voltage $V$,
 one can observe a series of different patterns
 and finally reaches a regime of spatiotemporal chaos, or turbulence,
 called the dynamic scattering mode (DSM) \cite{deGennes.Prost-Book1995}.
There are two turbulent regimes here, called DSM1 and DSM2.
DSM2 is observed under higher voltages and composed of topological defect lines
 as sketched in \figref{fig13}(a).
More precisely, on applying a sufficiently high electric field,
 one first observes the DSM1 state, but it is only metastable
 and eventually replaced by the stable DSM2 state.
Here, DSM2 nucleates randomly and takes over the metastable DSM1 state
 through a random growth process, as shown in \figref{fig13}(b,c).
It is this process that we focus on here,
 in the context of kinetic roughening of interfaces.
Observations suggest that topological defects
 are stretched, multiplied
 (through broken alignment anchoring at the glass surfaces),
 and randomly transported by local turbulent flow,
 which result in random growth of DSM2 in macroscopic scales.
The interactions are expected to be short-ranged,
 because DSM2 is a regime of spatiotemporal chaos%
\footnote{
One can roughly estimate the Reynolds number for the DSM turbulence
 at $\re \approx 10^{-6} \ll 1$,
 using tens of micrometers as typical length scale,
 tens of micrometers per second as velocity scale,
 and known values of viscosities of nematic liquid crystal used here,
 namely MBBA \cite{deGennes.Prost-Book1995}.
Therefore, DSM2 is by no means a fully developed turbulence
 which would have long-ranged correlation.
},
 where correlation of flow fluctuations
 decays exponentially in space and time.

To work with the simplest situation of isotropic growth,
 Takeuchi and Sano \cite{Takeuchi.Sano-PRL2010,Takeuchi.etal-SR2011,Takeuchi.Sano-JSP2012} chose to use the so-called homeotropic alignment,
 where surface alignment of liquid crystal is perpendicular
 to the glass surfaces.
They also decided not to rely on spontaneous DSM2 nucleation,
 but rather to trigger it, by shooting ultraviolet laser pulses.
Figure~\ref{fig13}(b) shows a DSM2 interface,
 generated by pulses shot to a central point of the system,
 hence the interface is circular.
Moreover, by expanding the laser beam in one direction,
 they also shot pulses along a line, which then generates
 a flat interface (\figref{fig13}(c)).
This ability to change the interface geometry
 is important to test the theoretical hypothesis
 on the existence of different universality subclasses.
With the help of accurate temperature control,
 the authors were able to obtain 955 circular and 1128 flat interfaces
 under practically the same experimental conditions
 \cite{Takeuchi.Sano-JSP2012}.
Then they defined the height $h(x,t)$ as illustrated in \figref{fig13}(b,c).
For the circular case, $h(x,t)$ was chosen to be the local radius
 of the interface.
In this way, one can use data at all angular positions to evaluate
 statistical quantities, whereas the actual spatial coordinate
 is the azimuth $\theta$ (\figref{fig13}(b)).
In the following, however, $x$ is still used as the coordinate,
 which is formally converted by $x = \expct{h}\theta$.
Then it turned out that the Family-Vicsek scaling \pref{eq:FamilyVicsek}
 was accurately confirmed for both circular and flat interfaces,
 with the exponents $\alpha, \beta, z$ in agreement with the 1D KPZ class
\cite{Takeuchi.Sano-PRL2010,Takeuchi.etal-SR2011,Takeuchi.Sano-JSP2012}.

\subsection{Distribution}

\begin{figure}[t]
 \centering
 \includegraphics[clip]{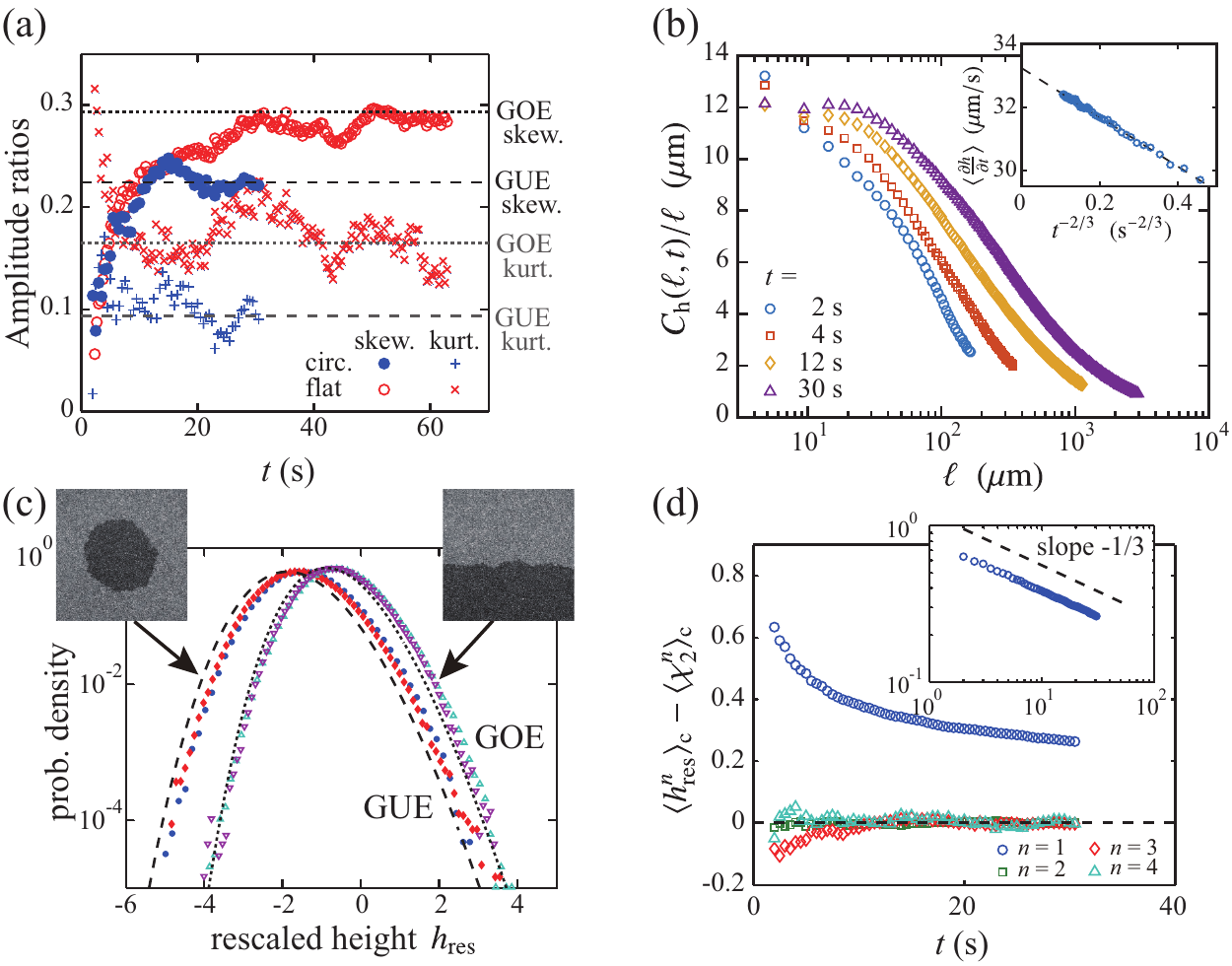}
 \caption{
Experimental results of the one-point distribution
 \cite{Takeuchi.etal-SR2011,Takeuchi.Sano-JSP2012}. 
(a) Skewness $\Sk(h)$ and kurtosis $\Ku(h)$ as functions of time.
(b) The height-difference correlation function $C_\mathrm{h}(\ell,t)$
 divided by $\ell$ (main panel) and the mean growth speed
 $\expct{\prt{h}{s}}$ (inset) in the circular case.
(c) Histograms of rescaled height $\hres$ for the circular case
 (solid symbols; $t = 10\unit{s}$ (blue $\bullet$)
 and $t=30\unit{s}$ (red $\blacklozenge$))
 and for the flat case (open symbols; $t = 20\unit{s}$ (turquoise $\triangle$)
 and $t = 60\unit{s}$ (purple $\triangledown$)).
Those are compared with the GUE and GOE Tracy-Widom distributions
 (dashed and dotted lines, respectively).
(d) Finite-time corrections in the $n$th-order cumulant $\hres$
 for the circular case.
The inset shows the correction for the mean in the log-log scales.
Reprints with permission from \cite{Takeuchi.Sano-JSP2012} (a,c)
 and from \cite{Takeuchi.etal-SR2011} (d) with some adaptations.
}
 \label{fig14}
\end{figure}%

With the data obtained thereby, the authors first measured
 the scaling exponents and found agreement with the 1D KPZ class.
Then they measured the skewness $\Sk(h) = \cum{h^3}/\cum{h^2}^{3/2}$
 and kurtosis $\Ku(h) = \cum{h^4}/\cum{h^2}^2$,
 which do not require estimation of the scaling coefficients.
Significant differences were found
 between the circular and flat interfaces,
 which were moreover in agreement with the values of the GUE and GOE
 Tracy-Widom distributions, respectively (\figref{fig14}(a)).
The circular and flat interfaces indeed show different fluctuation properties.

To make a more direct comparison with the Tracy-Widom distributions,
 the authors evaluated scaling coefficients by the method described
 in \secref{sec:Rescaling}.
By \eqref{eq:VinfEstim}, the parameter $v_\infty$
 was precisely estimated
 with four significant digits (\figref{fig14}(b) inset for the circular case).
Thanks to the rotational symmetry of the system, $\lambda = v_\infty$
 (\eqref{eq:ParamEstim3}).
In contrast, the parameter $A$ could be determined
 from the height-difference correlation function
 $C_\mathrm{h}(\ell,t) \simeq A\ell$ (\eqref{eq:ParamEstim1})
 only with two significant digits (\figref{fig14}(b) main panel),
 because of the rather narrow region
 in which the scaling \pref{eq:ParamEstim1} appears,
 as well as finite-time effect.
This restricted the precision of $\Gamma = A^2\lambda/2$ as well.
However, because it was already good enough to distinguish
 the GUE and GOE Tracy-Widom distributions, 
 the authors took the variance
 $\cum{h^2} \simeq (\Gamma t)^{2/3} \cum{\chi_i^2}$
 and used the variance of $\chi_2$ and $\chi_1$
 to improve the precision of $\Gamma$ \cite{Takeuchi.Sano-JSP2012}
 (see also footnote~\ref{ft:ParamEstim}).
The parameter $A$ was obtained in turn by $A = \sqrt{2\Gamma/\lambda}$.

With these scaling coefficients, the authors obtained
 the rescaled height $\hres$ by \eqref{eq:RescaledHeight}
 and constructed the histogram (\figref{fig14}(c)).
The results clearly confirmed the GUE and GOE Tracy-Widom distributions,
 apart from a small horizontal shift due to finite-time corrections.
Indeed, by plotting $\cum{\hres^n}-\cum{\chi_i^n}$
 with $i=2$ (circular) and $i=1$ (flat),
 the authors found that the mean ($n=1$)
 has a pronounced finite-time correction%
\footnote{
The presence of the correction $\sim t^{-1/3}$ in the mean $\expct{\hres}$
 can be easily understood as a contribution from the next leading term
 in \eqref{eq:Height}, which is $\mathcal{O}(t^0)$.
}
 that decays as $t^{-1/3}$, while the corrections in the other cumulants
 became small enough during the observation time.
In summary, the emergence of the Tracy-Widom distributions,
 as well as the split into the circular and flat subclasses,
 were clearly found in this experiment.
Those results turned out not to be special properties of integrable systems,
 but universal features of the KPZ class!

\subsection{Spatial correlation}

\begin{figure}[t]
 \centering
 \includegraphics[clip]{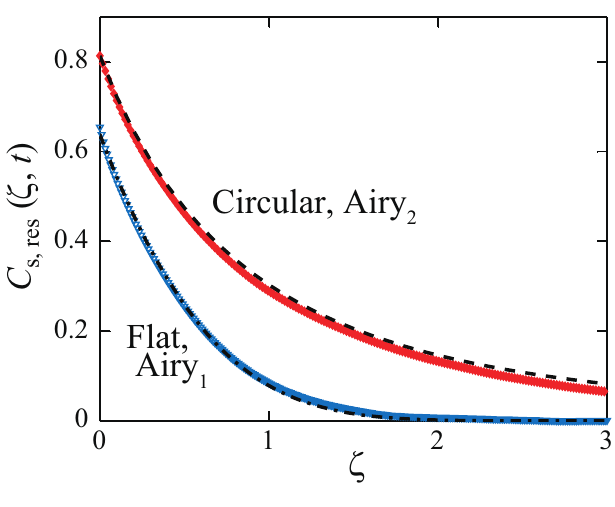}
 \caption{
Experimental results for the spatial correlation function
 \cite{Takeuchi.Sano-JSP2012}. 
The rescaled correlation function $C_\mathrm{s,res}(\zeta,t)$ 
 as defined in \eqref{eq:SpaceCorrRescaling} is shown
 against rescaled length $\zeta = (A\ell/2)(\Gamma t)^{-2/3}$,
 for the circular case (red solid symbols; $t=30\unit{s}$)
 and for the flat case (blue open symbols; $t=60\unit{s}$).
The data are compared with the two-point function
 of the Airy$_2$ correlation $g_2(\zeta)$ (dashed line)
 and the Airy$_1$ correlation $g_1(\zeta)$ (dashed-dotted line).
Reprints with permission from \cite{Takeuchi.Sano-JSP2012}
 with some adaptations.
}
 \label{fig15}
\end{figure}%

The scaling coefficients $\Gamma$ and $\lambda$ are also sufficient
 to test the predictions on the spatial correlation
 (\secref{sec:SpaceCorr}).
The authors measured the two-point spatial correlation function
 $C_\mathrm{s}(\ell,t)$ (\eqref{eq:SpaceCorr})
 and rescaled it according to \eqref{eq:SpaceCorrRescaling}.
The rescaled function $C_\mathrm{s,res}(\zeta,t)$ was then plotted
 against rescaled length $\zeta$ (\figref{fig15})
 and excellent agreement was found with the Airy$_2$ correlation $g_2(\zeta)$
 for the circular case%
\footnote{
Note that, thanks to the use of the rescaled radius for the circular case,
 one does not need to take into account the influence of the global curvature.
The term $-\zeta^2$ in \eqref{eq:AiryProcessConv} is omitted
 and one can directly compare with the function $g_2(\zeta)$.
}, and with the Airy$_1$ correlation $g_1(\zeta)$ for the flat case.
This allows us to infer that
 the interface profiles we can observe by microscope,
 i.e., \figref{fig13}(b,c), are nothing but realizations
 of the Airy$_2$ and Airy$_1$ processes
 (circular and flat interfaces, respectively) at large $t$.
More importantly, the agreement with them also implies that
 the qualitatively different decay of the spatial correlation function
 is also present in this experimental system
 (see discussions below \eqref{eq:SpaceCorrRescaling}).

\subsection{Temporal correlation}  \label{sec:ExpTimeCorr}

\begin{figure}[t]
 \centering
 \includegraphics[clip]{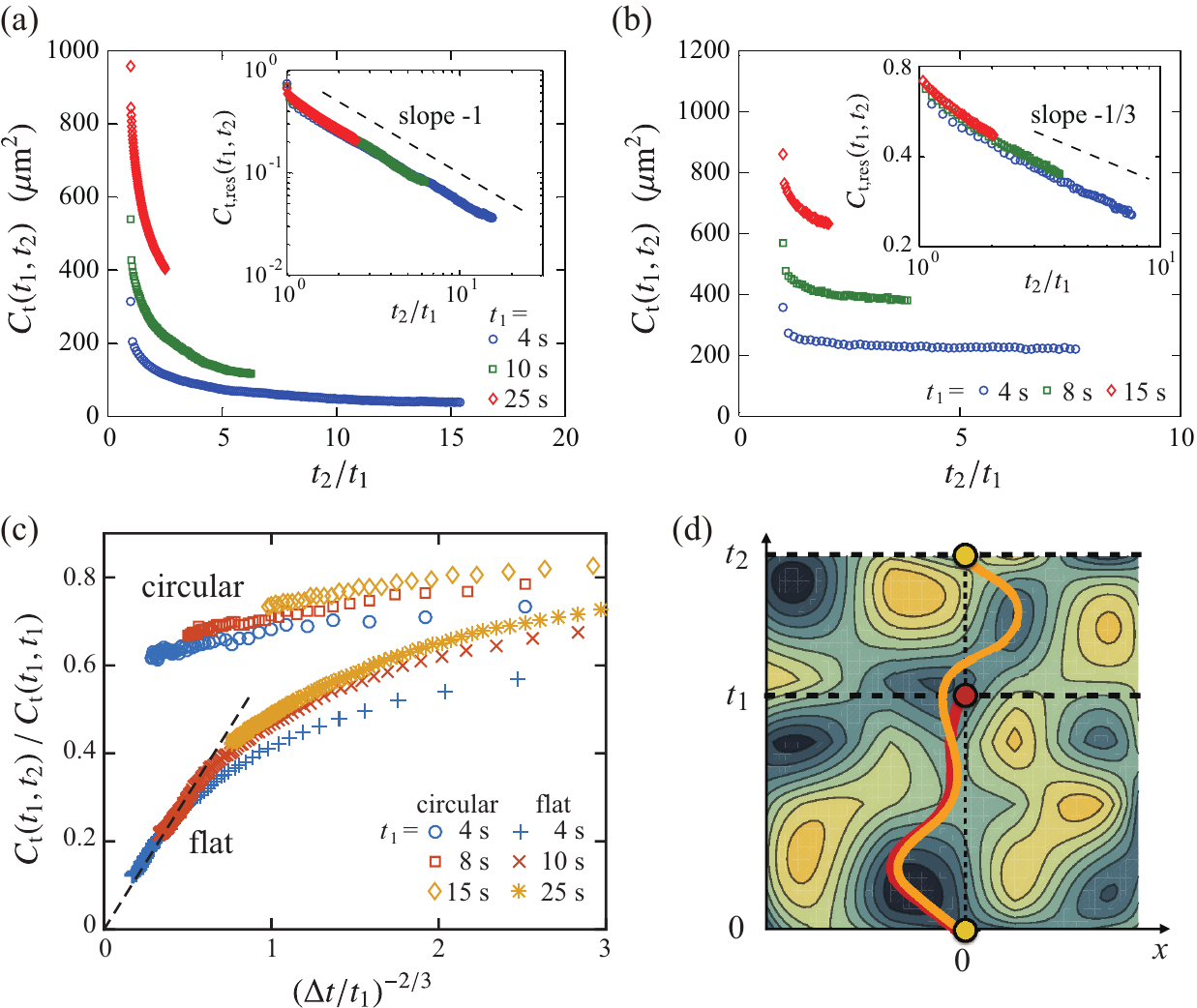}
 \caption{
Experimental results of the temporal correlation function
 \cite{Takeuchi.Sano-JSP2012,DeNardis.etal-PRL2017}.
(a,b) Two-time correlation function $C_\mathrm{t}(t_1,t_2)$
 against $t_2/t_1$ for the flat (a) and circular (b) interfaces.
The insets show the rescaled function $C_\mathrm{t,res}(t_1,t_2)$
 \cite{Takeuchi.Sano-JSP2012}.
(c) Ratio of the two-time correlation function,
 $C_\mathrm{t}(t_1,t_2) / C_\mathrm{t}(t_1,t_1)$,
 against $(\Delta t/t_1)^{-2/3}$ with $\Delta t \equiv t_2-t_1$.
The data for the circular interfaces clearly show that
 $C_\mathrm{t}(t_1,t_2)$ remain strictly positive
 in the limit $t_2 \to\infty$.
Dashed line is a guide for the eyes.
(d) Sketch of the two-time problem in terms of the directed polymer
 under random potential \cite{DeNardis.etal-PRL2017}.
Reprints with permission from \cite{Takeuchi.Sano-JSP2012} (a,b)
 and from \cite{DeNardis.etal-PRL2017} (d) with some adaptations.
}
 \label{fig16}
\end{figure}%

In contrast to the theoretically well-understood spatial correlation,
 temporal correlation has remained
 a very difficult property to study analytically.
However, experimentally, it is one of the natural and straightforward
 quantities to measure.
It is quantified here by the two-time correlation function%
\footnote{
Note that for the circular case the azimuth $\theta$ was actually fixed
 in \eqref{eq:TimeCorrDef}.
}
\begin{equation}
 C_\mathrm{t}(t_1,t_2) \equiv \cum{h(x,t_1)h(x,t_2)}.  \label{eq:TimeCorrDef}
\end{equation}
According to the usual scaling ansatz, it is expected to scale as
$C_\mathrm{t}(t_1,t_2) \simeq (\Gamma^2 t_1 t_2)^{1/3} F_\mathrm{t}(t_2/t_1)$
 with a scaling function $F_\mathrm{t}(\cdot)$.
Therefore, one can define the rescaled two-time function by
\begin{equation}
 C_\mathrm{t,res}(t_1,t_2)
 \equiv (\Gamma^2 t_1 t_2)^{-1/3} C_\mathrm{t}(t_1,t_2)
 \simeq F_\mathrm{t}(t_2/t_1).  \label{eq:TimeCorrRescaling}
\end{equation}

Figures~\ref{fig16}(a,b) show the experimental results
 for the flat (a) and circular (b) interfaces.
The scaling \pref{eq:TimeCorrRescaling} is confirmed in the insets,
 despite somewhat larger finite-time correction for the circular case.
Here we notice an important difference between the two cases:
 when $t_2$ is increased with fixed $t_1$,
 the correlation $C_\mathrm{t}(t_1,t_2)$ decays as usually expected
 in the flat case (\figref{fig16}(a)),
 but in the circular case it decays very slowly and remains large
 during the observation time (\figref{fig16}(b)).
Indeed, while the rescaled function decays
 as $F_\mathrm{t}(\tau) \sim \tau^{-\bar\lambda}$ with $\bar\lambda = 1$
 for the flat case,
 in agreement with earlier numerical simulations \cite{Kallabis.Krug-EL1999},
 for the circular case $\bar\lambda$ was found to be $1/3$.
Kallabis and Krug \cite{Kallabis.Krug-EL1999} conjectured that the relation
 obtained for the flat linear growth equations (such as the EW equation),
 $\bar\lambda = \beta + d/z$ (here $d=1$), also holds in nonlinear growth
 such as the KPZ class.
However, the experimental results for the circular case
 suggest a modification of the conjecture as follows%
\footnote{
For the circular case, the circumference
 (or equivalently the effective system size)
 grows with $t$, faster than the correlation length does,
 $\xi(t) \sim t^{1/z}$.
One can therefore argue that the fluctuations at different spatial positions
 do not influence the fluctuation at a given point in later times,
 asymptotically.
This is an intuitive reason why, in \eqref{eq:KallabisKrug},
 the term $d/z$ is not expected for the circular case.
}
 \cite{Takeuchi.Sano-JSP2012}:
\begin{equation}
 \bar\lambda = \begin{cases} \beta + d/z & \text{(flat subclass)}, \\ \beta & \text{(circular subclass)}. \end{cases}  \label{eq:KallabisKrug}
\end{equation}
Moreover, the exponent $\bar\lambda = 1/3$ for the circular case suggests that,
 when translated back into $C_\mathrm{t}(t_1,t_2)$
 by \eqref{eq:TimeCorrRescaling}, $C_\mathrm{t}(t_1,t_2)$
 remains strictly positive even if $t_2 \to \infty$.
In other words, positive fluctuations tend to remain positive forever
 and vice versa, suggesting a sort of ergodicity breaking in the circular case.
This can be clearly seen
 by plotting $C_\mathrm{t}(t_1,t_2) / C_\mathrm{t}(t_1,t_1)$ against
 $(\Delta t/t_1)^{-2/3}$ with $\Delta t \equiv t_2-t_1$
 \cite{DeNardis.etal-PRL2017} (\figref{fig16}(c)):
 here, while the flat data go to $0$ as $t_2 \to \infty$ linearly in this plot,
 the circular data clearly indicate a strictly positive asymptotic value.
The authors also measured persistence properties of the height fluctuations,
 or more precisely, the probability $P_\text{pers}(t_1,t_2)$ that
 $\delta h(x,t) = h(x,t) - \expct{h(x,t)}$ with fixed $x$
 does not change sign between times $t_1$ and $t_2$
 \cite{Takeuchi.Sano-JSP2012}.
The results showed power-law decay in both cases,
 $P_\text{pers}(t_1,t_2) \sim (t_2/t_1)^{-\theta_\pm}$
 where $\pm$ indicates the sign of the fluctuations,
 with the following exponent values:
\begin{equation}
 \begin{cases} \theta_+ = 1.35(5) \\ \theta_- = 1.85(10) \end{cases} \text{(flat)} \qquad \text{and} \qquad \begin{cases} \theta_+ = 0.81(2) \\ \theta_- = 0.80(2) \end{cases} \text{(circular)}.   \label{eq:ExpTheta}
\end{equation}
Here, the numbers in the parentheses
 indicate the range of error in the last digit(s).
The results for the flat case are consistent with earlier numerics
 \cite{Kallabis.Krug-EL1999} and the inequality $\theta_+ \neq \theta_-$
 was associated with the KPZ nonlinear term%
\footnote{
Therefore, the values of $\theta_+$ and $\theta_-$ are exchanged
 if $\lambda < 0$ \cite{Kallabis.Krug-EL1999}.
}
 $(\lambda/2)(\vnabla h)^2$.
In contrast, this asymmetry seems to vanish in the circular case;
 this remains theoretically unexplained so far.
Moreover, since the mean persistence time, i.e.,
 mean time needed for the fluctuations to change the sign,
 is given by $\int P_\text{pers}(t_1,t_2) \rd t_2$,
 $\theta_\pm < 1$ for the circular case implies that it diverges,
 hence ergodicity breaking in this case.

Recently, theoretical approaches have also become capable of dealing with
 some two-time properties.
Ferrari and Spohn \cite{Ferrari.Spohn-SIG2016}
 considered TASEP and derived the exponent $\bar\lambda$
 for the circular, flat, and stationary subclasses.
The exponents were indeed $\bar\lambda = 1/3$ and $1$
 for the circular and flat cases, respectively,
 accounting for the experimental observations.
They also showed that the two-time correlation
 for the stationary subclass, when appropriately rescaled,
 coincides with that of the fractional Brownian motion
 with Hurst exponent $1/3$ \cite{Ferrari.Spohn-SIG2016};
 specifically,
\begin{equation}
 F_\mathrm{t}(\tau) \sim \cum{\chi_0^2} \frac{\tau^{-1/3}}{2} \[ 1 + \tau^{2/3} - (\tau -1)^{2/3} \] \qquad \text{(stationary subclass)},
\end{equation}
 even though the process $\chi(x,t)$ in time is \textit{not}
 the fractional Brownian motion,
 because its one-time distribution is not Gaussian
 but the Baik-Rains distribution.
Furthermore, very recently, even analytic expressions
 of the two-time joint distribution for the circular case
 were obtained by different authors \cite{Dotsenko-JSM2013,Johansson-CMP2017,DeNardis.etal-PRL2017,Nardis.Doussal-JSM2017}.
While the formulas in \cite{Dotsenko-JSM2013,Johansson-CMP2017}
 remain difficult to evaluate numerically,
 de Nardis and Le Doussal \cite{DeNardis.etal-PRL2017,Nardis.Doussal-JSM2017}
 found that an easily tractable form can be obtained
 by the replica Bethe ansatz approach to the Lieb-Liniger model
 \pref{eq:LL}, i.e., the KPZ equation.
Their derivation was based on a hint that
 directed polymers in the point-to-point problem
 tend to overlap near the origin $(x,t)=(0,0)$ (\figref{fig16}(d)),
 more and more if $h(0,t_1)$ is larger, since then
 the random potential landscape tends to have a deeper path
 between times $0$ and $t_1$ (recall that $h(0,t) = \log Z(0,t)$
 is the (minus) free energy of the directed polymer).
This led them to focus on such eigenstates of $\hat{H}_\mathrm{LL}$
 that all particles are bounded during this time period,
 and successfully obtained an approximate generating function
 for the two-time fluctuations. 
The deviation from the full problem turned out to be superexponentially small,
 in the order of $\mathcal{O}(\e^{-\frac{8}{3}\hres(0,t_1)^{3/2}})$.
With this generating function,
 a number of quantities were numerically evaluated,
 such as the cumulants and the two-time correlation function
 conditioned for large $h(x,t_1)$,
 which were found in agreement experimentally and numerically
 \cite{DeNardis.etal-PRL2017}.
In particular, their theory accounted for the strictly positive
 two-time correlation $C_\mathrm{t}(t_1,t_2) > 0 ~(t_2\to\infty)$
 in the circular case, in terms of the overlap of directed polymers
 (\figref{fig16}(d)).

\subsection{Toward stationary subclass and general initial conditions}

While the liquid-crystal experiment has successfully identified
 a number of universal features of the circular and flat subclasses,
 the stationary subclass remains to be directly identified.
However, signatures of the stationary subclass, in particular
 the Baik-Rains distribution, were found
 by crossover analysis \cite{Takeuchi-PRL2013}
 and by power spectrum of height fluctuations $\delta h(x,t)$
 \cite{Takeuchi-JPA2017}.
The crossover toward the Baik-Rains distribution was also identified
 numerically \cite{Takeuchi-PRL2013,HalpinHealy.Lin-PRE2014}
 and theoretically \cite{DeNardis.etal-PRL2017,Nardis.Doussal-JSM2017}.

It is also interesting to explore other possible geometries of interfaces.
To this end, Fukai and Takeuchi \cite{Fukai.Takeuchi-PRL2017}
 developed a technique to generate an arbitrary initial condition
 in this liquid crystal system and studied interfaces
 growing from a ring of a finite radius, in the inward direction.
This is a circular problem with curvature
 opposite to the usual circular subclass, and interestingly,
 they found that the statistical properties agreed
 with those of the \textit{flat} subclass, until the interfaces
 become too small in circumference and eventually collapse
 at the center of the ring \cite{Fukai.Takeuchi-PRL2017}.
From a similar perspective, recently a number of theoretical studies
 \cite{Quastel.Remenik-a2016,Chhita.etal-a2016,Matetski.etal-a2017,LeDoussal-a2017} have also considered initial conditions
 more general than the standard ones.
It is hoped that such theoretical and experimental studies
 will lead to deeper understanding of the subclass structure
 of the KPZ universality class.

\section{Higher dimensions}  \label{sec:HigherDimension}

In Secs.\ref{sec:Exact1}-\ref{sec:Exp} we have essentially
 considered the 1D case, but the KPZ class is by no means restricted
 to the 1D problem.
There are no known integrable models
 that are able to provide exact results for higher dimensions,
 but numerical and experimental studies have been carried out to
 investigate the distribution and correlation functions
 for the 2D case, i.e.,
 surfaces growing in the three-dimensional space.
Interestingly, the subclass structure persists in higher dimensions.
More specifically, on the basis of numerical results by
 Halpin-Healy \cite{HalpinHealy-PRL2012,HalpinHealy-PRE2013}
 and Oliveira \etal \cite{Oliveira.etal-PRE2013},
 there are four canonical subclasses in 2D:
 sphere (point-to-point), cylinder (point-to-line), flat (point-to-surface),
 and stationary.
Experiments on 2D thin film growth
 \cite{HalpinHealy.Palasantzas-EL2014,Almeida.etal-PRB2014,Almeida.etal-EL2015}
 indeed found that the experimentally obtained histograms were consistent
 with the numerical one for the flat 2D subclass.

Theoretically, Kloss \etal \cite{Kloss.etal-PRE2012}
 employed a non-perturbative renormalization group
 approach and evaluated the two-point correlation function
 of the stationary subclass, $g_0(\zeta)$.
They found, for 1D, a remarkable agreement
 with the exact result \cite{Prahofer.Spohn-JSP2004}.
For 2D and 3D, they provided an estimate of an amplitude ratio
 $R$ that contains $\cum{\chi_0^2}$, i.e.,
 an analogue of the Baik-Rains variance for each dimensionality.
Halpin-Healy's 2D numerics showed $R = 0.944 \pm 0.031$
 \cite{HalpinHealy-PRE2013,HalpinHealy-PRE2013a}
 in a remarkable agreement with Kloss \etal's estimate $R = 0.940$
 \cite{Kloss.etal-PRE2012}.
It remains a great challenge to invent a theoretical framework
 that can cope with other quantities and subclasses
 of the higher-dimensional KPZ class.

\section{Concluding remarks}  \label{sec:Perspectives}

The author hopes that these lecture notes provided
 a convincing overview that our understanding of the KPZ class,
 in particular for the 1D case, has significantly deepened
 by recent extensive investigations.
As a reminder, \tblref{tbl:subclass} gives a summary
 of the three subclasses established in the 1D KPZ class.
One might have the impression that
 most important pieces of work were already done.
However, if we recall the history of the equilibrium
 critical phenomena that is briefly overviewed in \secref{sec:Introduction},
 we notice that we are probably still in its mid 20th century or so,
 by which exact solutions of the 2D Ising model were found and
 related experimental observations were made.
Is there any fundamental theoretical framework,
 such as counterparts of Wilson's renormalization group
 and conformal field theory, hidden behind this development
 for non-equilibrium scale-invariant fluctuations?
Can we understand higher-dimensional problems equally well
 eventually, or at the level comparable to the recent developments
 on the 3D Ising model?
Somewhat optimistically, the author believes that
 there are fundamental open problems waiting to be found,
 in the future of studies on the KPZ class and related problems.

\section*{Acknowledgments}

The author is indebted to enlightening discussions
 with many theoreticians and mathematicians, in particular
 P. L. Ferrari, P. Grassberger, T. Halpin-Healy, P. Le Doussal, M. Pr\"ahofer,
 T. Sasamoto, G. Schehr, and H. Spohn,
 to learn the theoretical contents described in these lecture notes.
Among others, a series of lectures delivered by T. Sasamoto
 were important for the author to prepare many parts of the manuscript.
Of course, should any mistake be found in these lecture notes,
 it would be entirely the fault of the author.
The author acknowledges the theoretical curves of the Tracy-Widom distributions
 [\figref{fig14}(c)] provided by M. Pr\"ahofer \cite{TracyWidomTable},
 those of the Airy$_1$ and Airy$_2$ two-point correlation functions
 [\figref{fig15}] by F. Bornemann \cite{Bornemann-MC2010},
 permissions of figure reprints by M. A. C. Huergo [\figref{fig1}(a)],
 by P. Yunker [\figref{fig1}(b)],
 and by J. De Nardis and P. Le Doussal [\figref{fig16}(d)],
 and useful comments on the manuscript by T. Halpin-Healy.
Finally, the author appreciates financial support
 by KAKENHI from Japan Society for the Promotion of Science
 (No. JP25103004, JP16H04033, JP16K13846),
 the National Science Foundation under Grant No. NSF PHY11-25915,
 and 2016 Tokyo Tech Challenging Research Award.



\bibliographystyle{elsarticle-num} 
\bibliography{kpzlecnote}

%

\end{document}